\documentclass[preprintnumbers, aps, floatfix, onecolumn, preprintnumbers, letterpaper, superscriptaddress,nofootinbib,11pt]{revtex4}
\usepackage{dcolumn}
\usepackage{graphicx}
\usepackage{amsmath}
\usepackage{amsfonts}
\usepackage{amssymb}
\usepackage{microtype}
\usepackage{subfigure}
\usepackage{makeidx}
\usepackage{bm}
\usepackage{epsf}
\usepackage[colorlinks=true,citecolor=blue,linkcolor=black,urlcolor=black]{hyperref}
\usepackage{color}
\usepackage{multirow,dcolumn}
\usepackage{graphicx}
\usepackage{mathrsfs}
\graphicspath{{Images/}}

\begin{document}

\title{Exact charged black hole solutions in $D$-dimensions in $f(R)$ gravity}

\author{Zi-Yu Tang}
\email{tangziyu@sjtu.edu.cn}
\affiliation{Center for Gravitation and Cosmology, College of Physical Science
and Technology, Yangzhou University, Yangzhou 225009, China}
\affiliation{Center for Astronomy and Astrophysics, School of Physics and Astronomy,
Shanghai Jiao Tong University, Shanghai 200240, China}

\author{Bin Wang}
\email{wang\_b@sjtu.edu.cn  (Corresponding Author)}
\affiliation{Center for Gravitation and Cosmology, College of Physical Science
and Technology, Yangzhou University, Yangzhou 225009, China}
\affiliation{School of Aeronautics and Astronautics, Shanghai Jiao Tong University, Shanghai 200240, China}

\author{Eleftherios Papantonopoulos}
\email{lpapa@central.ntua.gr} \affiliation{Physics Division,
National Technical University of Athens, 15780 Zografou Campus,
Athens, Greece.}

\begin{abstract}
We consider Maxwell-$f(R)$ gravity and obtain an exact charged black hole solution with dynamic curvature in  $D$-dimensions. Considering a  spherically symmetric metric ansatz and without specifying the form of $f(R)$ we find a general black hole solution in $D$-dimensions. This general black hole solution can reduce to the Reissner-Nordstr\"om (RN) black hole in $D$-dimensions in Einstein gravity and to the known charged black hole solutions with constant curvature in $f(R)$ gravity. Restricting the parameters of the general solution we get polynomial solutions which reveal novel properties when compared to RN black holes. Specifically we study the solution in $(3+1)$-dimensions in which the form of $f(R)$ can be solved explicitly giving a dynamic curvature and compare it with the RN black hole.  We also carry out a detailed study of its thermodynamics.
\end{abstract}

\maketitle

\section{Introduction}

Modified theories of gravity with the presence of higher-order curvature terms have been introduced in a attempt to  describe the early and late cosmological history including the early time inflation and the late time acceleration. Another motivation to study such theories is for understanding the presence of dark matter and the confrontation between gravity theories and the recent  observations \cite{Nojiri:2006ri,Copeland:2006wr,Nojiri:2010wj,Clifton:2011jh}. On more theoretical grounds,  higher-order corrections to the Einstein-Hilbert term lead to a renormalizable and thus quantizable gravitational theory \cite{Stelle:1976gc}.  Therefore, modified theories of gravity with higher-order corrections, provide a deeper understanding of General Relativity (GR). The presence of high curvature correction terms in GR provides  some interesting physical results, for examples, it makes the condensation harder to be formed in holographic superconductivity \cite{Gregory:2009fj, Kuang:2013oqa}; it modifies the low-energy tensor perturbation spectrum in string backgrounds \cite{Gasperini:1997up}; and it influences the dynamics of stellar structure \cite{Hansraj:2020xmz}.

At very early times, to avoid restricting the gravitational Lagrangian to be only a linear function of R, variable modified theories of gravity that contain some of the four possible second-order curvature invariants were proposed with the effects of quadratic Lagrangians. Besides, one particular class of models that includes higher order curvature invariants as functions of the Ricci scalar is the $f(R)$ gravity model \cite{DeFelice:2010aj}-\cite{Vainio:2016qas}.  Although such theories exclude contributions from any curvature invariants other than $R$,  they could also avoid the Ostrogradski’s instability \cite{Ostrogradsky:1850fid} which proves to be problematic for general higher derivative theories \cite{Woodard:2006nt}. Considering the gravitational collapse, one would expect all the matter present to be absorbed  by the black hole, so the final state should be vacuum except for the presence of electromagnetic fields associated with the black hole. Therefore, it is of great interest to study the stationary black hole solutions in Maxwell-$f(R)$ theory, which describe how the nonlinearity of $f(R)$ guides and affects the contribution of Maxwell field on the geometry.

Although it is well known that the black hole solutions in GR are also solutions in many modified gravity theories, including a large class of $f(R)$ gravity \cite{Multamaki:2006zb}, the black hole solutions in modified gravity theories that differ from the solutions in GR can be used to distinguish different modified gravity theories or give the constraints on parameters of the models by gravitational waves or shadows. Therefore the searching for exact solutions in the $f(R)$ theory of gravity is important but challenging because the equations of motion are very complicated with higher order terms.

Nevertheless a lot of exact and numerical solutions are obtained by various of methods, including the method of Lagrange multiplier \cite{Sebastiani:2010kv} and the so-called generator method \cite{Amirabi:2015aya}. Firstly the most simple case is a special class of $f(R)$ gravity with constant curvature, the solutions of which (including Schwarzschild-like \cite{Multamaki:2006zb}, RN-like \cite{delaCruzDombriz:2009et,Moon:2011hq} and Kerr-Newman-like solutions \cite{Cembranos:2011sr}) differ from GR solutions only by a constant coefficient $f'(R_0)$ that however can be absorbed into the Newton’s constant. While the solutions with dynamic curvatures \cite{Multamaki:2006zb,Sebastiani:2010kv} have real distinctions with the solutions in GR. Besides, static spherically symmetric solutions with perfect fluid \cite{Multamaki:2006ym}, Yang-Mills field \cite{Mazharimousavi:2011bf}, nonlinear Yang-Mills field \cite{Moon:2011hq,Mazharimousavi:2011nc}, Maxwell field and non-linear electromagnetic fields \cite{HabibMazharimousavi:2011yj} \cite{Hollenstein:2008hp,Rodrigues:2015ayd,Hurtado:2020gic} are also obtained. By Noether symmetries, axially symmetric solutions can be derived  from exact spherically symmetric solutions \cite{Capozziello:2009jg}. In addition, an interesting correspondence between the solutions in Einstein-conformally invariant Maxwell theory and the solutions in $f(R)$ gravity without matter field in arbitrary dimensions is shown in \cite{Hendi:2009sw,Hendi:2011eg}. Furthermore, the spherically symmetric vacuum solutions in $f(R)$ gravity in higher dimensions were studied 
in \cite{Amirabi:2015aya,Carames:2009ek}.

In this work, considering a spherically symmetric metric ansatz with $g_{tt}g_{rr}=-1$, without specifying the form of the function $f(R)$, we obtain exact charged black hole solutions in general $D$-dimensional $f(R)$ gravity. Constraining the parameters of the general solution we get polynomial solutions which reveal some interesting properties when compared with RN black holes. We then focus on the $(3+1)$-dimensions, where the form of $f(R)$ can be solved explicitly from the polynomial solution and we discuss in details the thermodynamics of this solution studying the First-Law, entropy, Hawking temperature and heat capacity.

The work is organized as follows. In Section \ref{general sol} we discuss the general $D$-dimensional solution. In Section \ref{spec sol} we show that for $c_2=0$ the general solution can reduce to RN black hole in $D$-dimensions in Einstein gravity and to the known charged black hole solutions with constant curvature in $f(R)$ gravity. While for $c_1=-1$ we can obtain new charged $D$-dimensional solutions, which can reduce to the spherically symmetric vacuum solutions discussed in \cite{Amirabi:2015aya,Carames:2009ek}. In Section \ref{sol in dimen} we study the polynomial solution in $D$-dimensions. In Section \ref{scalar-tensor} we find the corresponding equivalent theories in scalar-tensor gravity. In Section \ref{spec31sol} we study the thermodynamics of the black hole solution in $(3+1)$-dimensions. Finally Section \ref{coclusion} are our conclusions.

\section{General Solution in $D$-dimensions ($D\ge 3$) }
\label{general sol}

We consider a $D$-dimensional action
\begin{eqnarray}
I=\int d^D x\sqrt{-g}\left[\frac{1}{2\kappa}\left(R+f(R)-2\Lambda\right)-\frac{1}{2}F_{\mu\nu}F^{\mu\nu}\right]~, \label{action}
\end{eqnarray}
which, except for the Ricci scalar $R$, includes a general function of $f(R)$ that is not specified, a Maxwell field and a cosmological constant $\Lambda$.

In this paper we use $\kappa=8\pi G=1$. By variation of the above action we  obtain the field equations
\begin{eqnarray}
    &&I_{\mu\nu}\equiv R_{\mu\nu}\left(1+f_R\right)-\frac{1}{2}g_{\mu\nu}\left(R+f(R)-2\Lambda\right)+\left(g_{\mu\nu}\square-\nabla_\mu \nabla_\nu\right)f_R-\kappa T_{\mu\nu} = 0~,\\
    &&\nabla_{\mu }F^{\mu \nu }=0~,
\end{eqnarray}
where
\begin{eqnarray}
    f_R &\equiv & \frac{df(R)}{dR}~,\\
    T_{\mu\nu}&=&-\frac{1}{2}g_{\mu\nu}F+2F_{\mu\sigma}{F_\nu}^\sigma~,\\
    F_{\mu\nu}&=&\partial_\mu A_\nu-\partial_\nu A_\mu~,\\
    A_a&=&h(r)(dt)_a~.
\end{eqnarray}

  We consider a metric ansatz with $g_{tt}g_{rr}=-1$ that contains only one unknown function $B(r)$
\begin{equation}
     ds^2=-B(r)dt^2+\frac{1}{B(r)}dr^2+r^2d\Omega_k^2~, \label{ansatz}
\end{equation}
where
\begin{equation}
d\Omega_k^2=\left\{
             \begin{array}{lr}
             d\theta_1^2+\sum\limits_{i=2}^{D-2}\prod\limits_{j=1}^{i-1}\sin^2 \theta_j d\theta_i^2 & k=1~, \\
             d\theta_1^2+\sinh^2{\theta_1}d\theta_2^2+\sinh^2{\theta_1}\sum\limits_{i=3}^{D-2}\prod\limits_{j=2}^{i-1}\sin^2 \theta_j d\theta_i^2 & k=-1~, \\
             \sum\limits_{i=1}^{D-2}d\theta_i^2 & k=0~, \\
             \end{array}
\right.
\end{equation}
represents the line element of a $(D-2)$-dimensional Einstein manifold with positive ($k=1$), negative ($k=-1$), or zero ($k=0$) curvature.

In GR, with or without the Maxwell field, assuming a general spherically symmetric metric ansatz, we can get the RN BH or Schwarzschild BH with the resultant relation $g_{tt} g_{rr}=-1$.  However, in general modified gravity theories the relation $g_{tt} g_{rr}=-1$ is not a necessary result and the resulting solutions are expected to be more complicated.  In $f(R)$ theories, it is difficult to get exact solutions for a general metric ansatz. Therefore, in order to get exact solutions  we assume this  special form of the metric. Another motivation to consider the specific form $g_{tt} g_{rr}=-1$ of the metric, is to be able to compare the resulting exact solutions with the  RN black holes in GR and study what is the effect of the $f(R)$ function on the known solutions in GR. In fact as we will discuss in the following content, in $(3+1)$-dimensions we obtained an exact charged  black hole with the  form $f(R)=2\alpha \sqrt{R-4\Lambda}$ of the $f(R)$ function and  we found some interesting properties when we compared them with RN black holes.

The non-zero components of the Einstein equation under the metric ansatz (\ref{ansatz}) are
\begin{eqnarray}
I_t^t&=&\frac{1}{4 r}\left[-2 r f_R(r) B''(r)-2 r B''(r)-2 (D-2) B'(r)\left(1+ f_R(r)\right)+2 r B'(r) f_R'(r)\right.\notag\\
&&\left.+4 (D-2) B(r) f_R'(r)+4 r B(r) f_R''(r)-2 r f(r)+r h'(r)^2+4 \Lambda  r-2 r R(r)\right]=0~,\\
I_r^r&=&\frac{1}{4 r}\left[-2 r f_R(r) B''(r)-2 r B''(r)-2 (D-2) B'(r)\left(1+ f_R(r)\right)+2 r B'(r) f_R'(r)\right.\notag\\
&&\left.+4 (D-2) B(r) f_R'(r)-2 r f(r)+r h'(r)^2+4 \Lambda  r-2 r R(r)\right]=0~,\\
I_{\theta_1}^{\theta_1}&=&-\frac{1}{4 r^2}\left[-4 r \left(r B'(r) f_R'(r)+B(r) \left((D-2) f_R'(r)+r f_R''(r)\right)\right)+2 r^2 (f(r)-2\Lambda +R(r))\right.\notag\\
&&\left.+4 (f_R(r)+1) \left(r B'(r)+(D-3) (B(r)-k)\right)+4 r B(r) f_R'(r)+r^2 h'(r)^2\right]=0~,\\
I_{\theta_i}^{\theta_i}&=&I_{\theta_1}^{\theta_1}~,
\end{eqnarray}
where $f(r) \equiv f\left(R(r)\right)$ and
\begin{equation}
R(r)=-\frac{r^2 B''(r)+2 (D-2) r B'(r)+(D-3) (D-2) (B(r)-k)}{r^2} \label{RB}
\end{equation}
is the Ricci scalar expressed  by the metric function $B(r)$ and its derivatives $B'(r)$ and $B''(r)$.

The equation $I_t^t-I_r^r=0$ can lead to a simple relation
\begin{equation}
B(r)f_R''(r)=0~,
\end{equation}
which gives
\begin{equation}
f_R(r)=c_1+c_2 r~, \label{fR}
\end{equation}
where $c_1$ and $c_2$ are integration constants. Since $f_R=f'(R)$ this relation becomes
\begin{eqnarray}
f'(R)&=&c_1+c_2 r(R)~,
\end{eqnarray}
which is solved as
\begin{eqnarray}
f(R)&=&c_1 R+c_2\int^R r(R)dR~.\label{central}
\end{eqnarray}
Now the integration constants $c_1, c_2$ in the rel. (\ref{central}) play the role of parameters of the function $f(R)$ as  $c_1$ is a dimensionless coefficient of the linear part of the function, while the parameter $c_2$ is the coefficient of the nonlinear part of the function that has dimension $\left[c_2\right]=L^{-1}$. The physical meaning of these parameters will be discussed in the following.

Therefore when
$c_1$ and $c_2$ are small this theory can be considered as a small perturbation of the Einstein gravity. The relation (\ref{central}) is one of the central relations in our work.

Besides, the $t$ component of the electromagnetic field equation gives
\begin{equation}
(D-2) h'(r)+r h''(r)=0~,
\end{equation}
from which we can solve
\begin{eqnarray}
h(r)=\left\{
             \begin{array}{lr}
             -\frac{q }{(D-3)r^{D-3}}+\phi_0 & D>3~,  \\
             q \ln{r}+\phi_0  & D=3~, \\
             \end{array}
\right.  \label{h(r)}
\end{eqnarray}
where $q$ and $\phi_0$ are integration  constants.

Using the eq. (\ref{h(r)}) we can find that the parameter $q$ is proportional to the charge $Q$
\begin{eqnarray}
Q&=&\frac{1}{4\pi}\int_S \ ^* F_{ab}=\frac{1}{8\pi}\int_S F^{ab}\varepsilon_{abc_1 c_2... c_{D-2}} \notag\\
&=&\frac{1}{4\pi} \int_S h'(r)\sqrt{-g}d\theta_1 d\theta_2...d\theta_{D-2}  \notag\\
&=&\frac{q}{4\pi} \omega_{D-2}~,
\end{eqnarray}
where $\omega_{D-2}$ is the volume of the unit $(D-2)$-dimensional spherical ($k=1$), hyperbolic ($k=-1$) and flat ($k=0$) horizons.
For $k=1$, we choose sphere $S^{D-2}$ topology while for $k=0$ we consider hypertorus $\mathbb{T}^{D-2}=\mathbb{R}^{D-2}/\mathbb{Z}^{D-2}$ then
\begin{eqnarray}
\omega_{D-2}=\left\{
             \begin{array}{lr}
             \frac{2\pi^{\frac{D-1}{2}}}{\Gamma\left(\frac{D-1}{2}\right)} & k=1~,  \\
             \left(2 \pi\right)^{D-2}  & k=0~.

             \end{array}
\right.
\end{eqnarray}
However for $k=-1$ there is a vast set of distinct compact manifolds that are difficult to calculate even for $D=4$ \cite{Ong:2015fha}.

Substituting the expressions of $f_R(r)$ and $h(r)$ into the Einstein equations, we have two independent equations  with two unknown functions $f(r)$ and $B(r)$,
\begin{eqnarray}
I_t^t&=& c_2 \left[B'(r)+\frac{(D-2) B(r)}{r}\right]- \frac{c_2}{2}  B'(r)-\frac{1}{2r}(c_1+c_2 r+1) \left[r B''(r)+(D-2) B'(r)\right]\notag\\
&&+\frac{q^2}{4 r^{2 (D-2)}}-\frac{1}{2}\left[f(r)-2\Lambda +R(r)\right]=0~,\\
 I_{\theta_1}^{\theta_1}&=&-\frac{1}{4 r^2}\left\{4 c_1 r B'(r)+4 r B'(r)+2 (c_1+1) (D-3) B(r)-4(D-3)\left( c_1+1+c_2 r\right)k \right.\notag\\
 &&\left.+\frac{q^2}{r^{2(D-3)}}+2 r^2 f(r)-4 \Lambda  r^2+2 r^2 R(r)\right\}=0~,
\end{eqnarray}
which contain both $D>3$ and $D=3$ cases.
We solve $f(r)$ from $I_{\theta_1}^{\theta_1}=0$,
\begin{eqnarray}
f(r)&=&\frac{2 \left\{(D-3) \left[-(c_1+1) B(r)+\left(c_1+c_2 r+1\right)k\right]
-(c_1+1) r B'(r)\right\}}{r^2}\\ \nonumber
&-&\frac{q^2}{2 r^{2 (D-2)}}+2\Lambda -R(r)~,\label{fBR}
\end{eqnarray}
and put it back into $I_t^t=0$, then a simple second order differential equation with respect to the metric function $B(r)$ is obtained
\begin{eqnarray}
&&(c_1+c_2 r+1) \left[r^2 B''(r)+2 (D-3)k\right]+r B'(r) \left[(c_1+1) (D-4)+c_2 r(D-3)\right]\notag\\
&&-2 B(r) \left[(c_1+1) (D-3)+c_2 r(D-2)\right]=q^2 r^{2(3-D)}~, \label{Metric equation}
\end{eqnarray}
from which we get  the general exact solution in $D$-dimensions ($D>3$) for the metric function $B(r)$
\begin{eqnarray}
B(r)&=&-\left\{3Dr^6\Gamma (D+2)\left(c_1+1\right)^2\left[-q^2 \left(2 D^2-7 D+6\right) r^{-2 D} \Gamma (2 D-2) _2F_1^*\left(1,2 (D-1);2 D-1;-\frac{c_1+1}{r c_2}\right)\right.  \right.\notag \\
&&+c_2^4 \ln \left|\frac{c_2 r+c_1+1}{c_2 r}\right| \left(2 (D-3) k \left(c_2 D r+c_1 (D-2)-3 c_2 r+D-2\right)+q^2 (D-2) r^{6-2 D}\right)\notag \\
&&\left. 2 r^{-8}(3-2 D)^2 (D-3) k \Gamma (D+1)\, _2 F_1\left(1,D+1;D+2;-\frac{c_1+1}{r c_2}\right) \right]\notag \\
&&-\left[(D-3) r^D \left(2 k r^D \left(c_2 D r+c_1 (D-2)-3 c_2 r+D-2\right)+c_4 (D-2) r^3\right)+q^2 (D-2) r^6\right] \notag \\
&&3 \left(c_1+1\right){}^2 D (2 D-3) r^{-2 D} \Gamma (D+1) \, _2F_1\left(1,D+1;D+2;-\frac{c_1+1}{r c_2}\right)\notag \\
&&+r c_2 \left[D \left(2 D^2-9 D+9\right) k \Gamma (D+2) \left(-2 \left(c_1+1\right)^2 (D-2)-6 c_2^2 r^2+3 c_2 \left(c_1+1\right) r\right)\right. \notag \\
&&-\left(c_1+1\right) (D+1) \Gamma (D+1)\left[3 c_2 D \left(2 D^2-9 D+9\right) r \left((3-D) k-c_3 (D-2) r^2\right) \right.\notag \\
&&\left.\left.+(2-D) r^{-2 D} \left(\left(2 D^2-9 D+9\right) r^D \left(2 \left(c_1+1\right) D k r^D+3 c_4 r^3\right)+3 q^2 D r^6\right)\right)\right]  \notag \\
&& \left. \right\} \left/ \left[3 \left(c_1+1\right) c_2^2 (D-3) (D-2) D (D+1) (2 D-3) r^2 \Gamma (D+1)\right]\right. , \label{General Solution}
\end{eqnarray}
where $c_3$ and $c_4$ are constants of integration, while $_2F_1$ and $_2F_1^*$ are the hypergeometric function and the regularized generalized hypergeometric function respectively.

Note that when we solve eq. (\ref{General Solution}), with the dimension $D$ to be  an arbitrary parameter, the domain  $|-\frac{c_1+1}{r c_2}|<1$ of the hypergeometric functions and the regularized generalized hypergeometric functions which appear in the general solution, is hard to be satisfied. But in fact, the dimension $D>3$ being  an integer,  the hypergeometric functions reduce to the logarithmic functions, whose domain $|\frac{c_2 r+c_1+1}{c_2 r}|>0$ can be satisfied with $r>0$ and $c_2>0$ avoiding the divergence $r=0$ and the zero $r=-\frac{c_1+1}{c_2}$. Note that the condition $|c_1|<<1$ has been used to ensure the deviation of our theory from Einstein gravity which is not large. Besides, the solution contains $(1+c_1)$ and $c_2$ in the denominator, therefore this general solution is only valid for $c_1\neq -1$ and $c_2\neq 0$. We will elaborate these special cases in the next section. 

In the action (\ref{action}) the presence of an explicit cosmological constant $\Lambda$ introduces a scale $1/L^2$ in the theory. However in (\ref{General Solution}) this
cosmological constant $\Lambda$ does not appear explicitly, though it shows
up in the field equations. This is because the function $f(R)$, rather than $R$, introduces another length scale which redefines the original cosmological constant to an effective cosmological constant $\Lambda_{eff}$. In the following parts, we will show that even in higher dimensional cases, the effective cosmological constant can be defined as the coefficient of
$r^2$ term in the metric function redefining the length scale of the theory.

For $D=3$ we obtain a new exact charged black hole solution
\begin{eqnarray}
B(r)&=&-\frac{1}{4 \left(c_1+1\right){}^3}\left\{\left(c_1+1\right) \left[q^2 \left(-4 c_2 r+c_1+1\right)-4 \left(c_1+1\right){}^2 c_3 r^2\right] \right.\notag \\
&&+2 \left(c_1+1\right)q^2 \left(-2 c_2 r+c_1+1\right) \ln{r}+4 c_2^2 q^2 r^2 Li_2\left(-\frac{r c_2}{c_1+1}\right)\notag \\
&&+2 c_4 \left[2 c_2^2 r^2 \ln{\left|\frac{c_2 r+c_1+1}{r}\right|}+\left(c_1+1\right) \left(-2 c_2 r+c_1+1\right)\right] \notag \\
&&\left.-2 c_2^2 q^2 r^2 \ln^2{r} +4 c_2^2 q^2 r^2 \ln{r} \ln{\left|\frac{c_2 r+c_1+1}{c_1+1}\right|}\right\}~, \label{General Solution3}
\end{eqnarray}
where 
\begin{equation}
    Li_2(z)=-\int_0^z \frac{\ln{\left(1-u\right)}}{u}du, \quad z\leq 1~,
\end{equation}
is the Spence's function or dilogarithm, a particular case of the polylogarithm (here we only consider the field of real number). To make sure the Spence's function $Li_2$ is valid for all the positive $r$, the condition $c_2>0$ is required when $|c_1|<<1$.

The cosmological constant $\Lambda$ does not show up in the solution, but we can know if the space is flat, AdS or dS by analysing its  asymptotic behaviors at spacial  infinity and origin
\begin{eqnarray}
&&B\left(r\to \infty \right) \to -Sgn\left(\Lambda_{\text{eff}}\right)\infty~,  \\
&&B\left(r \to 0\right)=-\frac{2 q^2 \ln (r)+q^2+2 c_4}{4 c_1+4} \to Sgn\left(1+c_1\right)\infty~,
\end{eqnarray}
where
\begin{eqnarray}
\Lambda_{\text{eff}}=-6 c_3-\frac{c_2^2 }{\left(c_1+1\right){}^3}\left[q^2 \left(3 \ln ^2\left(\frac{c_2}{c_1+1}\right)+\pi ^2\right)-6 c_4 \ln \left(c_2\right)\right].
\end{eqnarray}

In $(2+1)$-dimensions, near horizon solutions, asymptotically Lifshitz black hole solutions and rotating black holes with
exponential form of $f(R)$ theory have been discussed in \cite{Hendi:2014wsa}. They first gave the basic field equations as same as our equations with $D=3$, but their solution is different with our solution.

\section{Special Solutions}
\label{spec sol}

We will first consider solutions with $c_2=0$. In this case   $f'(R)=c_1$ and the eq. (\ref{Metric equation}) becomes
\begin{eqnarray}
(c_1+1) \left(r^2 B''(r)+(D-4) r B'(r)-2 (D-3) B(r)+2 D k-6 k\right)=q^2 r^{6-2 D}~,\label{Metric equation 2}
\end{eqnarray}
the solutions of which are
\begin{eqnarray}
B(r)=\left\{
             \begin{array}{lr}
             \frac{q^2 }{2 (c_1+1) (D-3) (D-2)r^{2(D-3)}}+c_3 r^2+c_4 r^{3-D}+k & D>3~,  \\
             c_4+\frac{c_3 r^2}{2}-\frac{q^2 }{2 (c_1+1)}\ln \left(2 (c_1+1) r\right)   & D=3~, \\
             \end{array}
\right.  \label{Solutionc2}
\end{eqnarray}
and then the functions $R(r)$ and $f(r)$ become
\begin{eqnarray}
R(r)&=&\left\{
             \begin{array}{lr}
             -c_3 (D-1) D-\frac{q^2 (D-4) }{2 (c_1+1) (D-2)r^{2 D-4}} & D>3~,  \\
             \frac{q^2}{2 (c_1+1) r^2}-3 c_3  & D=3~, \\
             \end{array}
\right. \label{R(r)} \\
f(r)&=&\left\{
             \begin{array}{lr}
             2\Lambda-c_3 (D-1) (2 c_1-D+2)-\frac{c_1 q^2 (D-4)}{2 (c_1+1) (D-2) r^{2 D-4}} & D>3~,  \\
             2\Lambda-2 c_1 c_3+c_3+\frac{c_1 q^2}{2 (c_1+1) r^2}  & D=3~. \\
             \end{array}
\right. \label{f(r)}
\end{eqnarray}

For $c_2=0$ we have two cases:

\subsection{ Non-constant $R$ }

For non-constant $R$, using the the solutions (\ref{R(r)})(\ref{f(r)}) we have for  $f(R)$
\begin{eqnarray}
f(R)&=&\left\{
             \begin{array}{lr}
             c_1 R+c_3(c_1+1) (D-1)(D-2)+2\Lambda & D>3~,  \\
             c_1 R+c_3\left(1+c_1\right)+2\Lambda  & D=3~, \\
             \end{array}
\right.
\end{eqnarray}
and then to have $f(R)=c_1R$ we get for the  parameter $c_3$ ($[c_3]=[\Lambda]=L^{-2}$)
\begin{eqnarray}
c_3=\left\{
             \begin{array}{lr}
             -\frac{2\Lambda }{(c_1+1)(D-1)(D-2)} & D>3~,  \\
             -\frac{2\Lambda }{c_1+1}  & D=3~. \\
             \end{array}
\right.
\end{eqnarray}
Then the metric function and the curvature function become
\begin{eqnarray}
B(r)&=&\left\{
             \begin{array}{lr}
             \frac{q^2 }{2 (c_1+1) (D-3) (D-2)r^{2(D-3)}}-\frac{2\Lambda  r^2}{(c_1+1) (D-2) (D-1)}+\frac{c_4}{r^{D-3}}+k & D>3~,  \\
             c_4-\frac{\Lambda  r^2}{ (c_1+1)}-\frac{q^2 }{2 (c_1+1)}\ln \left(2 (c_1+1) r\right)   & D=3~, \\
             \end{array}
\right. \label{Metric2=0}\\
R(r)&=&\left\{
             \begin{array}{lr}
             \frac{2D \Lambda }{(c_1+1) (D-2)}-\frac{q^2 (D-4) }{2 (c_1+1) (D-2)r^{2 D-4}} & D>3~,  \\
             \frac{q^2}{2 (c_1+1) r^2}+\frac{6 \Lambda }{c_1+1}  & D=3~. \\
             \end{array}
\right.
\end{eqnarray}

This result implies that it reduces to the Einstein gravity $R+f(R)-2\Lambda=(1+c_1)R-2\Lambda=R-2\Lambda$ when $c_1=0$. In this case the solutions (\ref{Metric2=0}) become
\begin{eqnarray}
B(r)&=&\left\{
             \begin{array}{lr}
             \frac{q^2 }{2  (D-3) (D-2)r^{2(D-3)}}-\frac{2\Lambda  r^2}{ (D-2) (D-1)}+\frac{c_4}{r^{D-3}}+k & D>3~,  \\
             c_4-\Lambda  r^2-\frac{q^2 }{2}\ln \left(2 r\right)   & D=3~, \\
             \end{array}
\right. \\
R(r)&=&\left\{
             \begin{array}{lr}
             \frac{2D \Lambda }{ (D-2)}-\frac{q^2 (D-4) }{2 (D-2)r^{2 D-4}} & D>3~,  \\
             \frac{q^2}{2 r^2}+6 \Lambda & D=3~, \\
             \end{array}
\right.
\end{eqnarray}
which after some parameterization
\begin{eqnarray}
\left\{
             \begin{array}{lr}
             c_4=-m_1,\quad q^2=2(D-3)(D-2)q_1^2 & D>3~,  \\
             c_4=-m_2+2q_2^2 \ln{2\ell},\quad q^2=4q_2^2 & D=3~, \\
             \end{array}
\right.
\end{eqnarray}
are exactly the standard higher-dimensional charged black hole solutions \cite{Chabab:2016cem} and the charged BTZ black hole solution \cite{Banados:1992wn} in Einstein-Maxwell theory
\begin{equation}
B(r)=\left\{
             \begin{array}{lr}
             k-\frac{2\Lambda}{(D-1)(D-2)}r^2 -\frac{m_1}{r^{D-3}}+\frac{q_1^2}{r^{2(D-3)}} & D>3~,  \\
             \frac{r^2}{l^2}-m_2-2 q_2^2 \ln{\frac{r}{\ell}}   & D=3~, \\
             \end{array}
\right.
\end{equation}
with dynamic curvatures
\begin{eqnarray}
R(r)&=&\left\{
             \begin{array}{lr}
             \frac{ 2D \Lambda }{D-2}-\frac{(D-3)(D-4) q_1^2}{r^{2(D-2)}} & D>3~,  \\
             \frac{2q_2^2}{ r^2}-\frac{6}{\ell^2}  & D=3~, \\
             \end{array}
\right.
\end{eqnarray}
where parameters $q_1$ and $m_1$ are related to the electric charge and the ADM mass of the BH, and $\ell$ is the AdS radius  $\Lambda=-\frac{1}{\ell^2}$.

\subsection{ Constant Curvature $R=R_0$}

For constant curvature $R=R_0$, we can take the trace of the Einstein equation and obtain
\begin{eqnarray}
R_0\left(1+f'(R_0)\right)-\frac{D}{2}\left(R_0+f(R_0)-2\Lambda\right)=\kappa T=0~,
\end{eqnarray}
where Maxwell electromagnetic field is traceless $T=g^{\mu\nu}T_{\mu\nu}=0$.
This relation can lead to
\begin{equation}
f(R_0)=-R_0+2\Lambda+c_0 R_0^{D/2}~,\label{f(R0)}
\end{equation}
which means $R_0+f(R_0)-2\Lambda=c_0 R_0^{D/2}$.

To see this consider the solutions eq. (\ref{R(r)}). We can see that for constant curvature the parameters must satisfy $q=0$ or $D=4$. The former condition $q=0$ leads to
\begin{eqnarray}
B(r)&=&k-\frac{R_0 r^2}{(D-1)D} +\frac{c_4}{r^{D-3}}~,\\
f(R_0)&=&2\Lambda+\frac{(2 c_1+2-D)R_0}{D}~,
\end{eqnarray}
which hold for all dimensions $D\geq 3$. Comparing with the eq. (\ref{f(R0)}), we can obtain the relation between the parameter $c_1$ and the constant curvature $R_0$
\begin{eqnarray}
c_1=\frac{c_0 D}{2}R_0^{\frac{D-2}{2}}-1~.\label{c1}
\end{eqnarray}

These geometries are exactly the same with the Schwarzschild black hole solutions in $D$-dimensions and the BTZ black hole solution in $3$-dimensions,
\begin{eqnarray}
B(r)=\left\{
             \begin{array}{lr}
             k-\frac{2\Lambda}{(D-1)(D-2)}r^2 -\frac{m}{r^{D-3}} & D>3~,  \\
             \frac{r^2}{l^2}-m  & D=3~. \\
             \end{array}
\right.
\end{eqnarray}

Specially in $4$-dimensions, the Ricci scalar is always a constant even with nonzero $q$,
\begin{eqnarray}
B(r)&=&k+\frac{q^2 }{4 (c_1+1)  r^{2}}-\frac{R_0}{12} r^2+\frac{c_4}{r}~,\\
f(R_0)&=&2\Lambda+ \frac{(c_1-1)R_0}{2}~,
\end{eqnarray}
where $c_1$ has the same relation as in eq. (\ref{c1}).

Note that $f'(R_0)=c_1$, after parametrizations we have
\begin{eqnarray}
B(r)&=&k-\frac{m}{r}+\frac{q^2 }{4 \left(f'(R_0)+1\right)  r^{2}}-\frac{R_0}{12} r^2,
\end{eqnarray}
which has been studied in \cite{Moon:2011hq,Soroushfar:2016nbu}. However this kind of solutions can not be distinguished with the RN black holes in Einstein gravity, since we can always adjust the gravitational constant to make them the same.

It is worth noticing that this charged solution with constant curvature only exists in $4$-dimensions while the Schwarzschild solution can be present in any higher dimensions.

\subsection{New charged $D$-dimensional solutions for $c_1=-1$}

We consider the special solutions for $c_1=-1$. In this case the linear term $(1+c_1)R$ in the action will disappear and the nonlinear term dominates. The eq. (\ref{Metric equation}) becomes
\begin{equation}
    c_2 r \left(r^2 B''(r)+(D-3) r B'(r)-2 (D-2) B(r)+2 D k-6 k\right)=q^2 r^{6-2 D}~,
\end{equation}
the solutions of which are
\begin{eqnarray}
B(r)=\left\{
             \begin{array}{lr}
             \frac{(D-3) k}{D-2}+\frac{q^2}{c_2 \left(2 D^2-9 D+9\right) r^{2 D-5}}+\frac{c_4}{r^{D-2}}+c_3 r^2 & D>3~,  \\
             c_3 r^2+\frac{c_4}{r}-\frac{q^2 }{9 c_2 r}\left(3 \ln (r)+1\right)   & D=3~, \\
             \end{array}
\right.
\end{eqnarray}
and then the functions $R(r)$ and $f(r)$ become
\begin{eqnarray}
R(r)&=&\left\{
             \begin{array}{lr}
             -\frac{(D-2) q^2 }{c_2 (2 D-3)r^{2 D-3}}-c_3 (D-1) D+\frac{(D-3) k}{r^2} & D>3~,  \\
             -\frac{q^2}{3 c_2 r^3}-6 c_3  & D=3~, \\
             \end{array}
\right. \\
f(r)&=&\left\{
             \begin{array}{lr}
             \frac{(D-3) k (2 c_2 r-1)}{r^2}+\frac{(D-2) q^2 }{c_2 (2 D-3)r^{2 D-3}}+c_3 (D-1) D-\frac{c_3^2}{2r^{2 D-4}}  +2 \Lambda & D>3~,  \\
             \frac{q^2}{3 c_2 r^3}-\frac{c_3^2}{2 r^2}+6 c_3+2 \Lambda  & D=3~. \\
             \end{array}
\right.
\end{eqnarray}

When $q=0$, these solutions are reduced to the solutions discussed in \cite{Amirabi:2015aya,Carames:2009ek}.

\section{Explicit solutions in various dimensions}

\label{sol in dimen}

In this Section we will discuss the forms and the properties of the general solution  (\ref{General Solution})  in various dimensions. First we discuss the simplest solution in $3$-dimensions.

\subsection{The solution in (2+1)-dimensions ($D=3$)}

We have found the solution for $D=3$  eq. (\ref{General Solution3}), to compare with the BTZ black hole we set $q=0$ and then the metric function becomes
\begin{eqnarray}
B(r)=c_3 r^2-\frac{c_4 \left(-2 c_2 r+c_1+1\right)}{2 \left(c_1+1\right){}^2}-\frac{c_2^2 c_4 r^2 }{\left(c_1+1\right){}^3}\ln \left|\frac{c_2 r+c_1+1}{r}\right|~.
\end{eqnarray}
From $\left|\frac{c_2 r+c_1+1}{r}\right|>0$ we can get $c_2(1+c_1)>0$.

The asymptotic behaviors of the metric function are
\begin{eqnarray}
B\left(r\to 0\right)&=&-\frac{c_4}{2(1+c_1)}\equiv B_0~,\\
B\left(r\to \infty \right)&=&\left(c_3-\frac{c_2^2 c_4}{\left(1+c_1\right)^3}\ln{c_2}\right)r^2+o\left(r^2\right)~,
\end{eqnarray}
where the leading order at $r\to \infty$ is $r^2$ term, so we define its coefficient as the effective cosmological constant
\begin{eqnarray}
\Lambda_{\text{eff}}=\frac{c_2^2 c_4}{\left(1+c_1\right)^3}\ln{c_2}-c_3~,
\end{eqnarray}
the sign of which can determine  the property of spacetime to be AdS or dS or if $\Lambda_{\text{eff}}=0$ to be flat.

To check the possibility of flat and dS black hole, we introduce a root function
\begin{eqnarray}
Root(r)=\frac{\left(c_1+1\right)}{2 c_2^2} \left(\frac{2 \left(c_1+1\right){}^2 c_3}{c_4}-\frac{-2 c_2 r+c_1+1}{r^2}\right)-\ln \left|\frac{c_2 r+c_1+1}{r}\right|~,
\end{eqnarray}
the roots of which are also the roots of the metric function $B(r)$. The derivative of $Root(r)$ is always positive under the condition $c_2(1+c_1)>0$
\begin{eqnarray}
Root'(r)=\frac{\left(c_1+1\right){}^3}{c_2^2 r^3 \left(c_2 r+c_1+1\right)}>0 \quad always,
\end{eqnarray}
and the asymptotic behaviors of root function at $r \to 0$ and $r\to \infty$ are respectively
\begin{eqnarray}
Root(r \to 0)&=&-\frac{\left(1+c_1\right)^2}{2c_2^2 r^2} \rightarrow -\infty,\\
Root(r\to \infty)&=&\frac{\left(c_1+1\right){}^3 c_3-c_2^2 c_4 \ln \left(c_2\right)}{c_2^2 c_4}=\frac{\left(1+c_1\right)^2\Lambda_{\text{eff}}}{2c_2^2 B_0}~,
\end{eqnarray}
where $\frac{\Lambda_{\text{eff}}}{B_0}>0$ indicates one horizon and $\frac{\Lambda_{\text{eff}}}{B_0} \leq 0$ indicates no horizon. If we want a dS black hole, there at least  two horizons exist, while if we want an AdS black hole, one horizon is required. It is clear that the solution we obtained can only represent AdS black hole spacetimes ($\Lambda<0$ and $B_0<0$) or pure dS spacetimes ($\Lambda>0$ and $B_0>0$).

\subsection{Solutions in $D$-dimensions ($D>3$)}
\label{SolutionsD}

The general solution (\ref{General Solution}) contains some special functions that are not easy to analyse. However, when we solve the eq.  (\ref{Metric equation}) in each dimension, the solutions become much simpler, only containing polynomials and logarithmic terms.
In the Appendix \ref{appendix1} we give  the solutions for $D=4,5,6$ dimensions. The solution  (\ref{General Solution4}) for $D=4$ has been discussed in \cite{Hollenstein:2008hp} while the solutions in higher dimensions are new and they have not been studied before.

To compare with the RN black holes and understand  the physical meaning of the constants of integration, we set the coefficients of the logarithmic terms to be zero and  then the solutions are just polynomials. Then from the solutions  (\ref{General Solution4}), (\ref{General Solution5}) and (\ref{General Solution6}) we obtain the general constraint for the parameters
\begin{eqnarray}
c_4(-1)^D (D-3) (D-2) (c_1+1)^{D-3} c_2^{D-3}+2 (D-3) k (c_1+1)^{2 D-5}+q^2 (D-2) c_2^{2 (D-3)}=0~, \label{constraint}
\end{eqnarray}
under which the  solutions (\ref{General Solution4}), (\ref{General Solution5}) and (\ref{General Solution6}) can be reduced to simpler polynomial solutions in $D$-dimensions
\begin{eqnarray}
B(r)&=&\frac{(D-3) k}{D-2}+\left[c_3 +\frac{2 c_2^2 (2 D-5) k}{(c_1+1)^2 (D-2)^2 (D-3)}\right]r^2+\sum _{n=1}^{D-3} \frac{2 k (-1)^{n+1} (c_1+1)^n}{(D-2) (n+2) c_2^n r^n} \notag \\
&&+\sum _{n=D-2}^{2 (D-3)} \frac{q^2 (-1)^n c_2^{2 D-n-6}}{(D-3) (n+2) r^n (c_1+1)^{2 D-n-5}}~.\label{simpler solution}
\end{eqnarray}

 Compared with RN black hole, our solution (\ref{simpler solution}) has more terms from $1/r$ to $1/r^{2(D-3)}$ while RN black hole only has two terms  $1/r$ and $1/r^{2(D-3)}$ as mass term and charge term respectively. Another interesting difference is that the constant term in our solution is a fraction $\frac{(D-3) k}{D-2}$ depending on dimension $D$, and this fraction can not be rescaled.   Moreover, the dynamic curvature $R(r)$ and the nonzero gravitational action $f(r)$ can not be simplified by the transformation of coordinates.

Firstly, the constant term of the solution is $\frac{D-2}{D-3}$, indicating that it is non-asymptotic flat with a deficit angle. This deficit angle will disappear at large $D$ limit, i.e. the constant term becomes $1$. Secondly, since the spacetime remains non-zero curvature even with zero mass, it can be interpreted as the global monopole solution in proper limits. Thirdly, there do exist some entanglements of mass and charge terms in the metric function. These entanglements become more and more complicated with the increase of the dimension, nevertheless we figure out the rules with any dimensions $D$.

Using relations (\ref{RB}) and (\ref{fBR}) we can obtain the expressions of $R(r)$ and $f(r)$
\begin{eqnarray}
R(r)&=&\frac{(D-3) k}{r^2}-(D-1)D\left[c_3  +\frac{2 c_2^2 (2D-5) k}{(c_1+1)^2 (D-3) (D-2)^2} \right]\notag\\
&&+\sum _{n=1}^{D-3} \frac{2 k (-1)^n (c_1+1)^n  \left(D^2-D (2 n+5)+n^2+5 n+6\right) }{(D-2) (n+2)c_2^{n} r^{n+2}} \notag \\
&&-\sum _{n=D-2}^{2 (D-3)} \frac{q^2 (-1)^n \left(D^2-D (2 n+5)+n^2+5 n+6\right)  }{(D-3) (n+2)(c_1+1)^{2 D-n-5} c_2^{n+6-2D}r^{n+2}}~,\\
f(r)&=&(D-1)(D-2-2 c_1)\left[c_3  +\frac{2 c_2^2 (2D-5) k}{(c_1+1)^2 (D-3) (D-2)^2}\right]+\frac{2 c_2 (D-3) k}{r}-\frac{q^2}{2 r^{2 D-4}} \notag\\
&&+\frac{(D-3) k (2 c_1-D+4)}{(D-2) r^2}-\sum _{n=1}^{D-3} \frac{2 k (-1)^n (c_1+1)^n  (-D+n+3)  (2 c_1-D+n+4)}{(D-2) (n+2)c_2^{n} r^{n+2}}\notag\\
&&+\sum _{n=D-2}^{2 (D-3)} \frac{q^2 (-1)^n  (-D+n+3) (2c_1-D+n+4)}{(D-3) (n+2)(c_1+1)^{2 D-n-5} c_2^{n+6-2D}r^{n+2}}+2\Lambda~,
\end{eqnarray}
and using the relation (\ref{fR}) we can get the exact expression of $f(R)$,
\begin{eqnarray}
f(R)&=&(D-1)(D-2-2 c_1)\left[c_3 +\frac{2 c_2^2 (2D-5) k}{(c_1+1)^2 (D-3) (D-2)^2}\right]+\frac{2 c_2^2 (D-3) k}{f'(R)-c_1}-\frac{q^2 c_2^{2 D-4}}{2 \left(f'(R)-c_1\right)^{2 D-4}} \notag\\
&&+\frac{(D-3) k (2 c_1-D+4)c_2^2}{(D-2) \left(f'(R)-c_1\right)^2}-\sum _{n=1}^{D-3} \frac{2 k (-1)^n (c_1+1)^n c_2^{2} (-D+n+3)  (2 c_1-D+n+4)}{(D-2) (n+2) \left(f'(R)-c_1\right)^{n+2}}\notag\\
&&+\sum _{n=D-2}^{2 (D-3)} \frac{q^2 (-1)^n c_2^{2D-4} (-D+n+3) (2 c_1-D+n+4) }{(D-3) (n+2)(c_1+1)^{2 D-n-5}\left(f'(R)-c_1\right)^{n+2}}+2\Lambda~.
\end{eqnarray}


 Defining an effective cosmological constant
\begin{eqnarray}
\Lambda_\text{eff}=-\frac{(D-1)(D-2)}{2}\left(c_1+1\right)\left[c_3 +\frac{2 c_2^2 (2 D-5) k}{(c_1+1)^2 (D-2)^2 (D-3)}\right]~,\label{Lambda}
\end{eqnarray}
we rewrite the functions $B(r)$, $R(r)$, $f(r)$, and $f(R)$ in the Appendix \ref{appendix2}.

\section{Equivalence to scalar-tensor theory}
\label{scalar-tensor}

\subsection{From $f(R)$ gravity to scalar-tensor theory}

For general $f(R)$ theory,
\begin{equation}
    S_{met}=\frac{1}{2\kappa}\int d^D x \sqrt{-g}f(R)+S_M\left(g_{\mu\nu},\psi\right),\label{f(R)}
\end{equation}
one can introduce a new field $\chi$ and write the dynamically equivalent action
\begin{equation}
    S_{met}=\frac{1}{2\kappa}\int d^D x \sqrt{-g}\left[f(\chi)+f'(\chi)\left(R-\chi\right)\right]+S_M\left(g_{\mu\nu},\psi\right).
\end{equation}
Variation with respect to $\chi$ leads to the equation
\begin{equation}
    f''(\chi)\left(R-\chi\right)=0.
\end{equation}
Therefore, $\chi=R$ if $f''(\chi)\neq 0$, which reproduces the action (\ref{f(R)}). Redefining the field $\chi$ by $\phi=f'(\chi)$ and setting
\begin{equation}
    V(\phi)=\chi(\phi)\phi-f\left(\chi(\phi)\right),
\end{equation}
the action takes the form
\begin{equation}
    S_{met}=\frac{1}{2\kappa}\int d^D x \sqrt{-g}\left[\phi R-V(\phi)\right]+S_M\left(g_{\mu\nu},\psi\right).\label{phiR}
\end{equation}

This is the Jordan frame representation of the action of a Brans-Dicke theory with Brans-Dicke parameter $\omega_0=0$.
It should be stressed that the scalar degree of freedom $\phi=f'(\chi)$ is quite different from a matter field, like all nonminimally coupled scalars, it can violate all of the energy conditions.

For the solutions in \ref{SolutionsD}, using
\begin{eqnarray}
\phi(r)&=&f_R(r)=c_1+c_2 r,\\
V(r)&=&\phi(r) R(r)-f(r),
\end{eqnarray}
we can obtain the potential as a function of $\phi$
\begin{eqnarray}
V(\phi)&=&V_0+V_1 \phi+\frac{q^2 c_2^{2 D-4}}{2 (\phi -c_1)^{2 D-4}}+\frac{c_2^2 (D-3) k (-2 c_1+(D-2) \phi +D-4)}{(D-2) (\phi -c_1)^2}-\frac{2 c_2^2 (D-3) k}{\phi -c_1}-2\Lambda \notag\\
&&-\sum _{n=D-2}^{2 (D-3)} \frac{q^2 (-1)^n c_2^{2 D-4} (-D+n+3) (c_1+1)^{-2 D+n+5} (-2 c_1+\phi  (-D+n+2)+D-n-4)}{(D-3) (n+2) (\phi -c_1)^{n+2}}\notag\\
&&+\sum _{n=1}^{D-3} \frac{2 c_2^2 k (-1)^n (c_1+1)^n (-D+n+3) (2 c_1+\phi  (-D+n+2)-D+n+4)}{(D-2) (n+2) (\phi -c_1)^{n+2}} .
\end{eqnarray}

Note that there are $(\phi - c_1)$ terms in the denominators, indicating the divergence of the potential at $\phi=c_1$. This  divergent point $\phi=c_1$ only appears at origin $r=0$, where a physical singularity exists. Also, the constant term of a scalar field   $\phi$ that comes from the transformation of $f(R)$ gravity  can always be rescaled. Besides, considering that the coefficient of the Einstein-Hilbert term in the action $(1+c_1)$ can be absorbed into the Newton's constant, we can always set $c_1=0$. Then the divergent point becomes $\phi=0$, which can be controlled in a potential.

Therefore though the $f(R)$ form is still unknown, but the equivalent scalar-tensor theory is clear.

\section{Explicit Solution in $(3+1)$-dimensions }
\label{spec31sol}

In  $(3+1)$-dimensions the form of $f(R)$ can be solved explicitly so we will discuss this solution in details and also study its thermodynamics.

The solution (\ref{simpler solution}) in $(3+1)$-dimensions writes
\begin{eqnarray}
B(r)&=&\frac{k}{2}+\frac{\left(c_1+1\right) k}{3 c_2 r}+\frac{q^2}{4 \left(c_1+1\right) r^2}+\frac{3 c_2^2 k+2 c_3 \left(c_1+1\right){}^2}{2 \left(c_1+1\right){}^2}r^2~,  \label{Solution4}\\
R(r)&=&-\frac{18 c_2^2 k}{(c_1+1)^2}-12 c_3+\frac{k}{r^2}~,\\
f(r)&=&\frac{c_1 k}{r^2}+\frac{2 c_2 k}{r}-\frac{9 \left(c_1-1\right) c_2^2 k}{\left(c_1+1\right){}^2}-6 \left(c_1-1\right) c_3+2\Lambda~,
\end{eqnarray}
from which we  can have an explicit  form of $f(R)$
\begin{eqnarray}
f(R)&=&\frac{2 c_2^2 k}{c_1}+c_1 \left(c_5+R\right)\pm 2 \sqrt{\frac{c_2^2 k W(R)}{c_1^2 \left(c_1+1\right){}^2}}~,\\
W(R)&=&-c_1 \left(5 c_2^2 k+6 c_3+2\Lambda \right)+c_1^2 \left(11 c_2^2 k-6 c_3+c_5-4 \Lambda +R\right)+2 c_2^2 k \notag\\
&&+c_1^3 \left(6 c_3+2 c_5-2\Lambda +2 R\right)+c_1^4 \left(6 c_3+c_5+R\right)~,
\end{eqnarray}
where the constant of integration $c_5$ should be set to
\begin{eqnarray}
c_5=-\frac{2c_2^2 k}{c_1^2}~,
\end{eqnarray}
to ensure that the only constant in the action is $-2\Lambda$.
After this constraint the $f(R)$ becomes
\begin{eqnarray}
f(R)=c_1 R\pm 2 \sqrt{\frac{c_2^2 k \left(9 \left(c_1-1\right) c_2^2 k+\left(c_1+1\right){}^2 \left(c_1 R-2\Lambda \right)+6 \left(c_1-1\right) c_3 \left(c_1+1\right){}^2\right)}{c_1 \left(c_1+1\right){}^2}}~.
\end{eqnarray}
Substituting $R(r)$ into this expression and comparing with $f(r)$, we finally get
\begin{eqnarray}
c_3=\frac{-2c_1 \Lambda -9 c_2^2 k-2\Lambda }{6 (c_1+1)^2}~,
\end{eqnarray}
 then the solution becomes
\begin{eqnarray}
B(r)&=&\frac{k}{2}+\frac{(c_1+1) k}{3 c_2 r}+\frac{q^2}{4 (c_1+1) r^2}-\frac{\Lambda  r^2}{3 (c_1+1)}~,\\
R(r)&=&\frac{4 \Lambda }{c_1+1}+\frac{k}{r^2}~,\\
f(r)&=&\frac{2k c_2}{r}+\frac{c_1 k}{r^2}+\frac{4 c_1 \Lambda }{c_1+1}~,\label{solutionc1c2}
\end{eqnarray}
with
\begin{eqnarray}
f(R)=c_1 R \pm 2 c_2 \sqrt{k\left(R-\frac{4\Lambda}{1+c_1}\right)}~.\label{action1}
\end{eqnarray}

We define the action of gravitational part as $F(R)=R+f(R)-2\Lambda$, then after a rescale we get,
\begin{eqnarray}
\frac{1}{2\kappa} F(R)&\equiv & \frac{1}{2\kappa}\left( R+f(R)-2\Lambda\right)\notag \\
&=&\frac{1}{2\kappa}\left((1+c_1) R\pm c_2\sqrt{ k  \left(R-\frac{4\Lambda}{1+c_1}\right)}-2\Lambda\right)\notag \\
&=&\frac{(1+c_1)}{2\kappa}\left(R\pm\frac{c_2}{1+c_1} \sqrt{ k  \left(R-\frac{4\Lambda}{1+c_1}\right)}-\frac{2\Lambda}{1+c_1}\right)\notag \\
&=&\frac{1}{2\kappa'} \left(R\pm\frac{c_2}{1+c_1} \sqrt{ k \left(R-4\Lambda'\right)}-2\Lambda'\right)~,
\end{eqnarray}
where $\Lambda'=\frac{\Lambda}{1+c_1}$ is an effective cosmological constant and $\kappa'=\frac{\kappa}{1+c_1}$. Then we can obtain the other equivalent form $\tilde{F}(R)$
\begin{eqnarray}
\tilde{F}(R)=R\pm\frac{c_2}{1+c_1} \sqrt{ k \left(R-4\Lambda'\right)}-2\Lambda'~.
\end{eqnarray}

In fact we can always  rescale to make $1+c_1=1$, then then we can adjust the parameter $c_2$ to change the relation between Einstein action term $R$ and the nonlinear action term $\sqrt{k\left(R-\frac{4\Lambda}{1+c_1}\right)}$.

Note that  when $c_1=0$ and $c_2=0$ the action (\ref{action1}) and the general solution will reduce to the Einstein gravity and RN black hole, but for this brunch with $c_2 \neq 0$, the solution can never reduce to the standard RN black hole solution even when $c_1=0$ and $c_2$ approaches to $0$. Here we give the solutions $B_0(r)$, $R_0(r)$, $f_0(r)$ and $F_0(r)$  to represent the solutions $B(r)$ , $R(r)$, $f(r)$ and $F(r)$ with $c_1=0$,
\begin{eqnarray}
B_0(r)&=&\frac{k}{2}+\frac{ k}{3 c_2 r}+\frac{q^2}{4  r^2}-\frac{\Lambda  r^2}{3 }~,\label{B0}\\
R_0(r)&=&4 \Lambda+\frac{k}{r^2}~,\label{R0}\\
f_0(r)&=&\frac{2k c_2}{r}~,\label{f0}\\
F_0(r)&=&2\Lambda+\frac{k}{r^2}+\frac{2k c_2}{r}~.\label{F0}
\end{eqnarray}

\begin{figure}[h]
\centering%
\subfigure[~$k=1$]{
 \includegraphics[width=.45\textwidth]{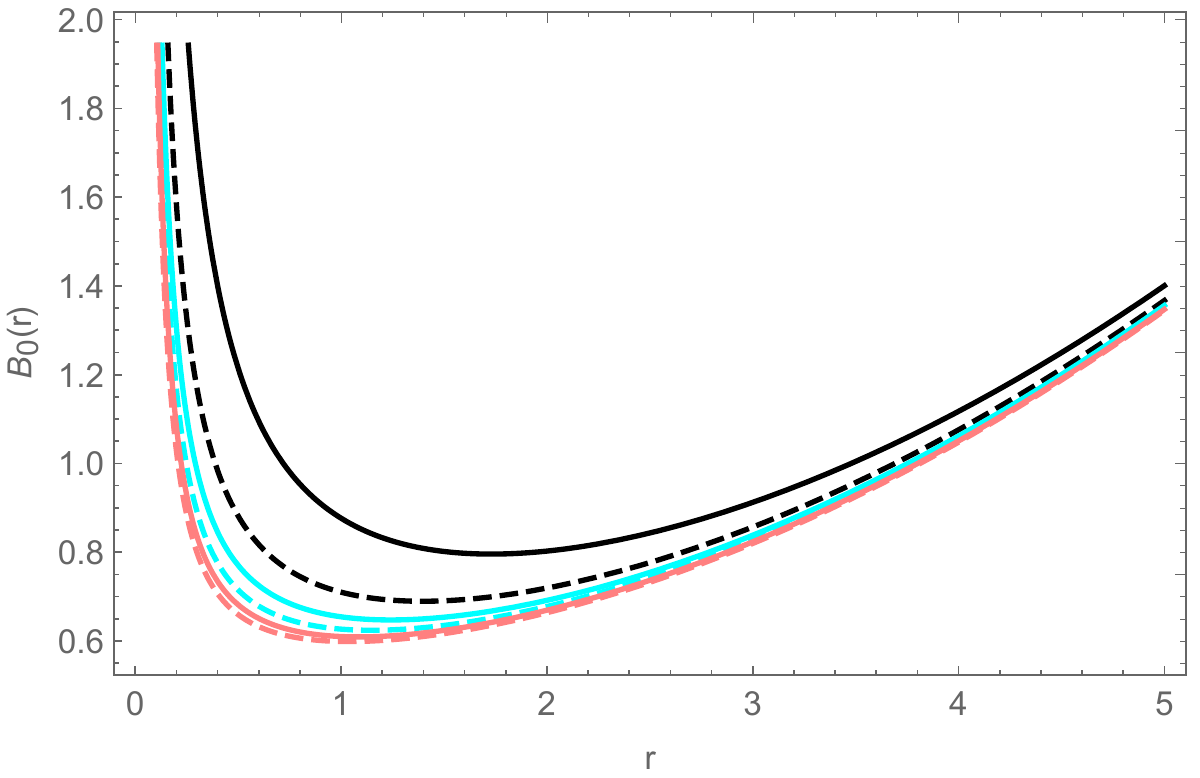} }
\subfigure[~$k=-1$]{
 \includegraphics[width=.45\textwidth]{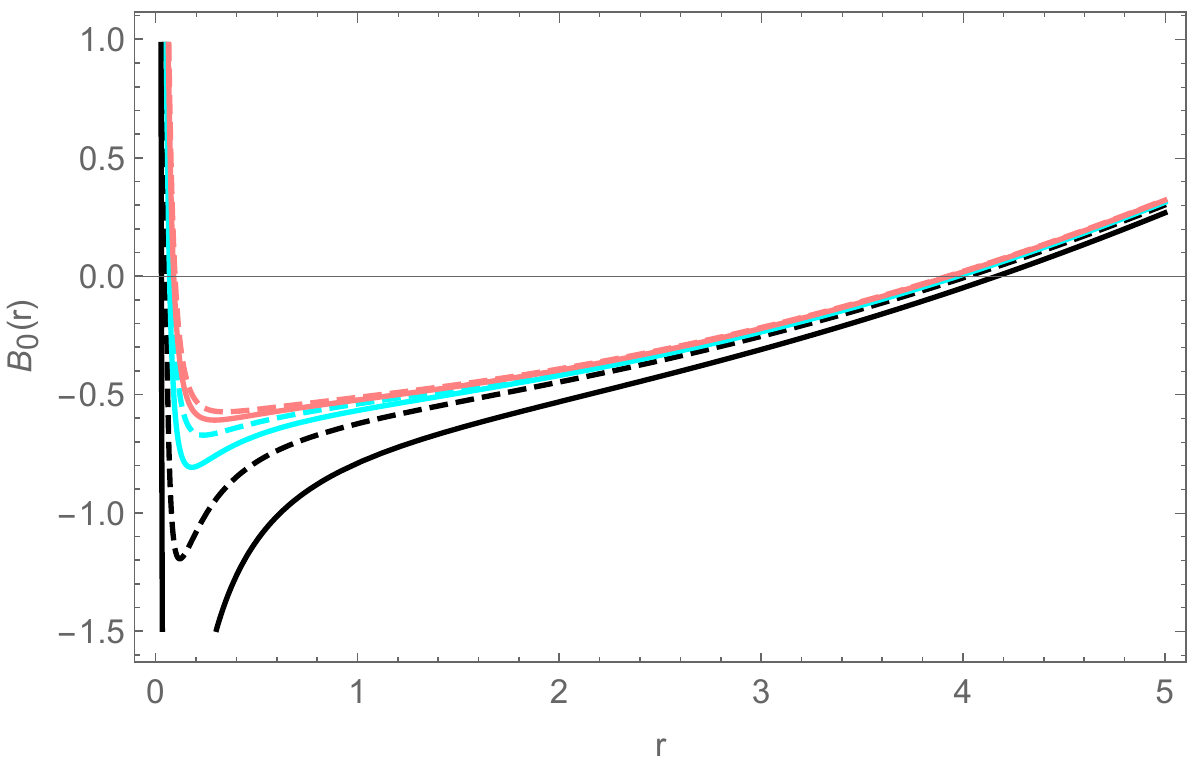} }
\caption{{}Metric functions $B_0(r)$ with $k=\pm 1$. The black curve, black dashed curve, cyan curve, cyan dashed curve, pink curve and pink dashed curve correspond to metric functions $B_0(r)$ with $c_2=1,2,3,4,5,6$ respectively. Other parameters are set as $q=0.4$ and $\Lambda=-0.1$. }
\label{fig:B0c2p}
\end{figure}

In the following figures we depict the plots of $B_0(r)$ to show the influence of parameter $c_2$. In Fig. \ref{fig:B0c2p}, the figures are plotted with positive $c_2$. It is clear that there can not exist any black holes for $k=1$, since all the terms in the metric function $B_0(r)$ are positive (Here we choose the asymptotic anti-de Sitter spacetimes). While for $k=-1$ black hole solutions exist and the black hole with larger $c_2$ has smaller radius of event horizon, which means that larger nonlinear gravitational action gives smaller black hole.

\begin{figure}[h]
\centering%
\subfigure[~$k=1$]{
 \includegraphics[width=.45\textwidth]{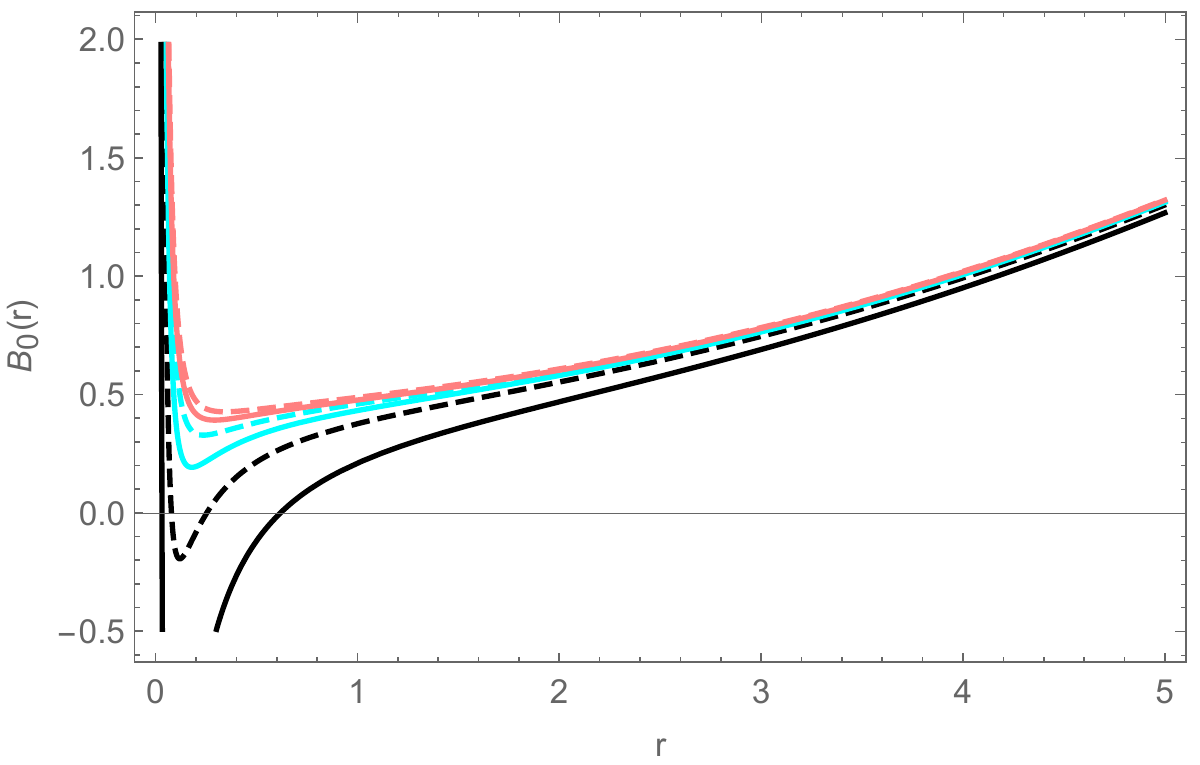} }
\subfigure[~$k=-1$]{
 \includegraphics[width=.45\textwidth]{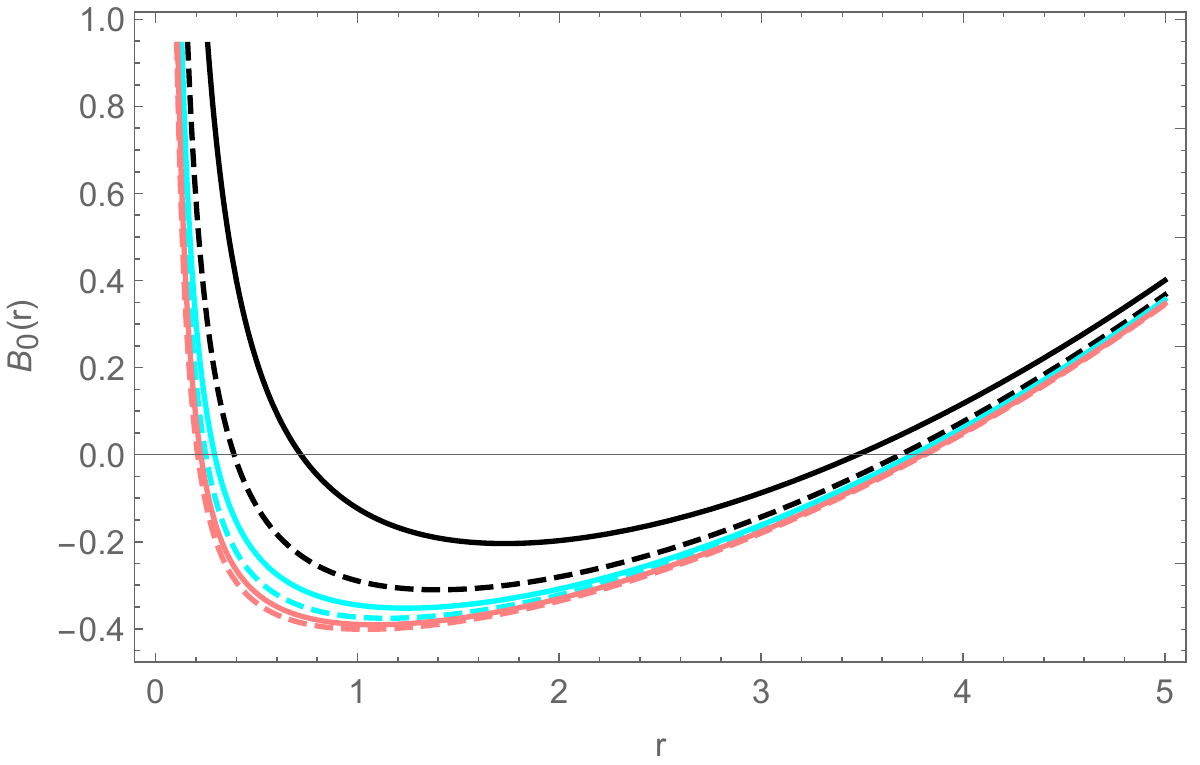} }
\caption{{}Metric functions $B_0(r)$ with $k=\pm 1$. The black curve, black dashed curve, cyan curve, cyan dashed curve, pink curve and pink dashed curve correspond to metric functions $B_0(r)$ with $c_2=-1,-2,-3,-4,-5,-6$ respectively. Other parameters are set as $q=0.4$ and $\Lambda=-0.1$. }
\label{fig:B0c2n}
\end{figure}

In Fig. \ref{fig:B0c2n}, the figures are plotted with negative $c_2$. It shows that for $k=1$ the black hole with larger absolute value of  $c_2$ has smaller radius of event horizon, which means that larger proportion of the nonlinear action part gives smaller black holes. While for $k=-1$ the black hole with larger absolute value of $c_2$ corresponds to larger radius of event horizon, indicating that larger proportion of the nonlinear action part gives larger black holes.

The above two figures are giving us some typical examples of the formation of black holes in (3+1)-dimensions due to the presence of the nonlinear term of the curvature.

Compared with RN black hole in $(3+1)$-dimensions,
\begin{eqnarray}
B_{\text{RN}}(r)&=& k -\frac{m_1}{r}+\frac{q_1^2}{r^{2}}-\frac{\Lambda}{3}r^2~,\\
R_{\text{RN}}(r)&=&2\Lambda~,\\
F_{\text{RN}}(r)&=&0~,
\end{eqnarray}
our solution has two terms related to the parameter $k$, which means the topology has more influence on the geometry. Besides, our solution also contains a dynamic curvature $R(r)$ and non-zero gravitational action $F(r)$, while the RN black hole has a constant curvature and zero gravitational action. Except the difference of the constant terms, we can rescale the parameters
\begin{eqnarray}
q=2q_1, \quad c_2=-\frac{k}{3m_1},
\end{eqnarray}
to make their metric functions very similar
\begin{eqnarray}
B_0(r)&=&\frac{k}{2}-\frac{ m_1}{ r}+\frac{q_1^2}{ r^2}-\frac{\Lambda  r^2}{3 }~,\\
B_{\text{RN}}(r)&=& k -\frac{m_1}{r}+\frac{q_1^2}{r^{2}}-\frac{\Lambda}{3}r^2~.
\end{eqnarray}



\begin{figure}[h]
\centering%
\subfigure[~$k=1$]{
 \includegraphics[width=.45\textwidth]{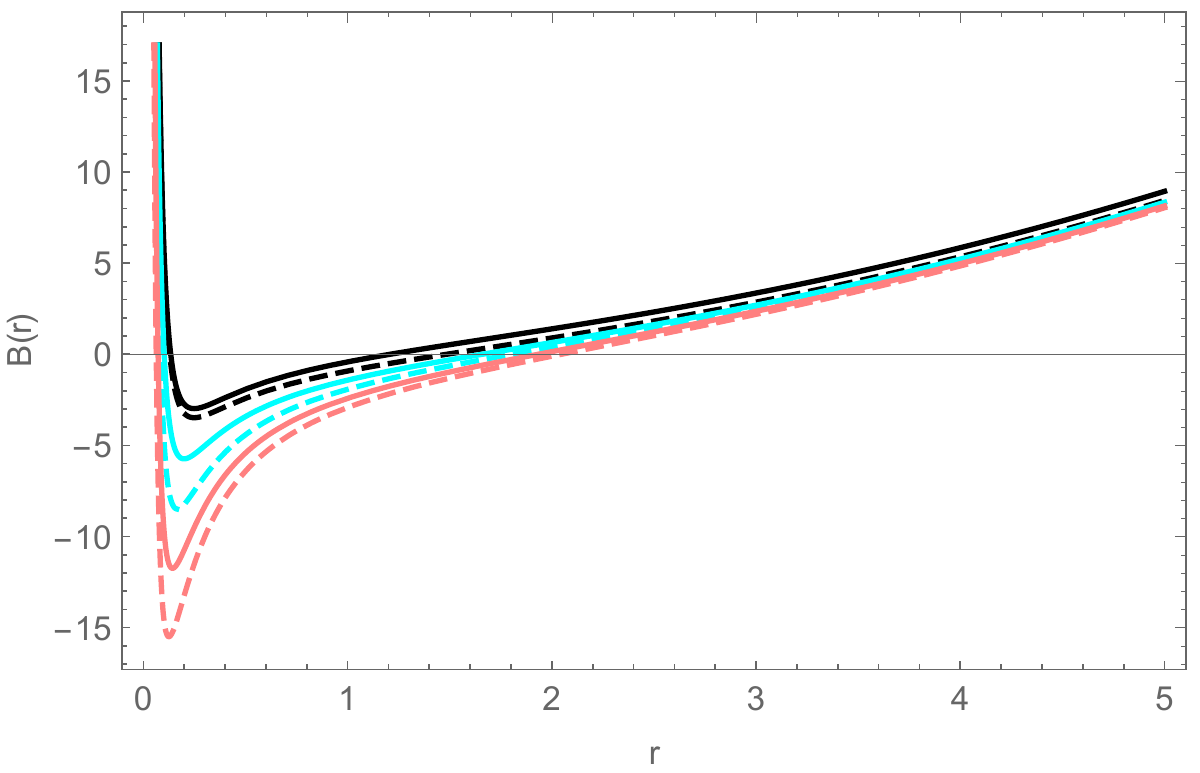} }
\subfigure[~$k=-1$]{
 \includegraphics[width=.45\textwidth]{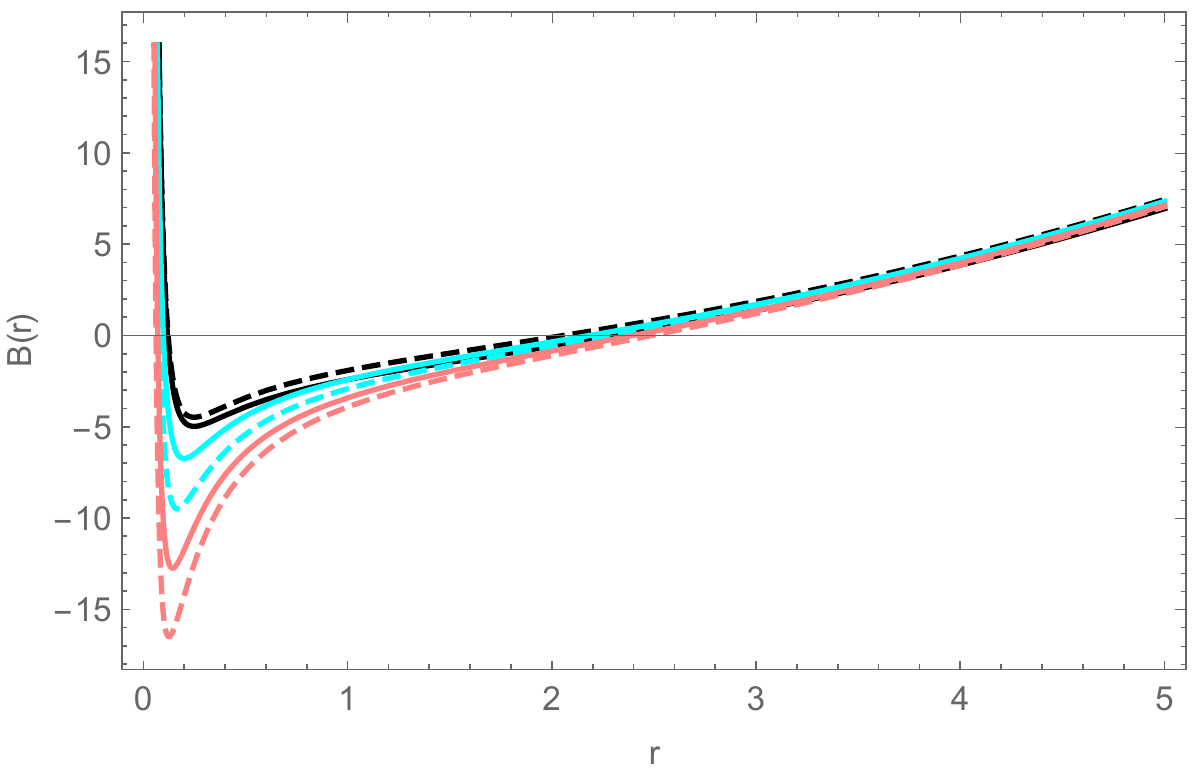} }
\caption{{}The black curve represents $B_{RN}(r)$ with $m_1=2$. While black dashed curve, cyan curve, cyan dashed curve, pink curve and pink dashed curve correspond to $B_0(r)$ with $m_1=2,2.5,3,3.5,4$ respectively. Other parameters are set as $q_1=0.5$ and $\Lambda=-1$.}
\label{fig:B0m1}
\end{figure}

In Fig. \ref{fig:B0m1}, we depict the plots of the metric functions $B_\text{RN}(r)$ and $B_0(r)$ with different parameters $m_1$ to observe the changes when the nonlinear term is present. Note that, no matter how we change the parameters, the metric functions $B_\text{RN}(r)$ and $B_0(r)$ always differ with a constant $\frac{k}{2}$, so we mainly plot the figures for $B_0(r)$. For black holes with the same charge and the cosmological constant, with the increase of $m_1$ (the nonlinear action part alleviates), the black holes have larger radius of event horizons and deeper depressions of geometry inside the event horizons.

Similar solution has been obtained in a recent paper \cite{Nashed:2019tuk} with an action of the form $F(R)=R-2\alpha\sqrt{R-8\Lambda}-\Lambda$.

\section{Thermodynamics}

In this section we will study the thermodynamics of the black hole solution in $(2+1)$-dimensions, including the First Law, Hawking temperature, entropy and heat capacity. Before we study the thermodynamics of our solution in $f(R)$ gravity, we first review the First Law for RN black hole in Einstein gravity to compare.

\subsection{RN black hole in Einstein gravity}

The metric of RN black hole in $D$-dimensions is
\begin{equation}
B(r)=1-\frac{m}{r^{D-3}}+\frac{q^2}{r^{2 (D-3)}},
\end{equation}
which gives the relation of the event horizon
\begin{equation}
m=q^2 r_+^{3-D}+r_+^{D-3}.
\end{equation}
Then the Hawking temperature and the entropy can be expressed by $r_+$
\begin{eqnarray}
T(r_+)&=&\frac{1}{4\pi}B'(r_+)=\frac{(D-3) r_+^{-2 D-1} \left(r_+^{2 D}-q^2 r_+^6\right)}{4 \pi },\\
S(r_+)&=&\frac{A(r_+)}{4G}=2\pi A(r_+)=\frac{4 \pi ^{\frac{D+1}{2}} r_+^{D-2}}{\Gamma \left(\frac{D-1}{2}\right)}.
\end{eqnarray}
For constant charge $Q$, the First Law $dM=TdS+\Phi_e dQ$ gives
\begin{eqnarray}
M=\int T(r_+)dS(r_+)=\frac{(D-2) \pi ^{\frac{D-1}{2}}\left(q^2 r_+^{3-D}+r_+^{D-3}\right)}{\Gamma \left(\frac{D-1}{2}\right)}=\frac{(D-2) \pi ^{\frac{D-1}{2}} m}{\Gamma \left(\frac{D-1}{2}\right)}.
\end{eqnarray}
Note that the expression of $r_+$ can be replaced by $m$, therefore the thermodynamic mass of the black hole can be described only by the parameter $m$.

\subsection{$f(R)$ black hole in $(3+1)$-dimensions}

In $(3+1)$-dimensions, the metric is
\begin{equation}
B(r)=\frac{1}{2}+\frac{1}{3 c_2 r}+\frac{q^2}{4 r^2},
\end{equation}
to compare with the RN black hole which can be written as
\begin{equation}
B(r)=\frac{1}{2}-\frac{m}{r}+\frac{q^2}{r^2},
\end{equation}
where $m=-1/(3c_2)$ and $q \to 2q$.


With the relation between $m$, $q$ and the event horizon $r_+$
\begin{equation}
m=\frac{q^2}{r_+}+\frac{r_+}{2},
\end{equation}
the Hawking temperature and Bekenstein-Hawking entropy in $f(R)$ gravity \cite{Akbar:2006mq, Camci:2020yre} become
\begin{eqnarray}
T(r_+)&=&\frac{r_+^2-2 q^2}{8 \pi  r_+^3},\\
S(r_+)&=&\frac{8 \pi ^2 r_+^2 \left(6 q^2+r_+^2\right)}{3 \left(2 q^2+r_+^2\right)}.
\end{eqnarray}
For constant charge $Q$, the First Law and heat capacity lead to
\begin{eqnarray}
M&=&\frac{2 \pi  r_+}{3}+\frac{8 \pi  q^2 r_+}{3 \left(2 q^2+r_+^2\right)}+\frac{4 \pi  q^2}{r_+}+\frac{8}{3} \pi  \sqrt{2} q \arctan\left(\frac{r_+}{\sqrt{2} q}\right),\\
C_Q&=&-\frac{16 \pi ^2 r_+^2 \left(r_+^2-2 q^2\right) \left(12 q^4+4 q^2 r_+^2+r_+^4\right)}{3 \left(r_+^2-6 q^2\right) \left(2 q^2+r_+^2\right)^2}, \label{C_Q}
\end{eqnarray}
where
\begin{equation}
r_+=m+\sqrt{m^2-2 q^2}.
\end{equation}

For neutral case, we have
\begin{equation}
M=\frac{2 \pi}{3}r_+=\frac{4\pi m}{3}, \quad C_Q=-\frac{16\pi^2 r_+^2}{3}=-\frac{64\pi ^2 m^2}{3}.
\end{equation}

\begin{figure}[h]
\centering%
\subfigure[~$T(r_+)$]{
 \includegraphics[width=.45\textwidth]{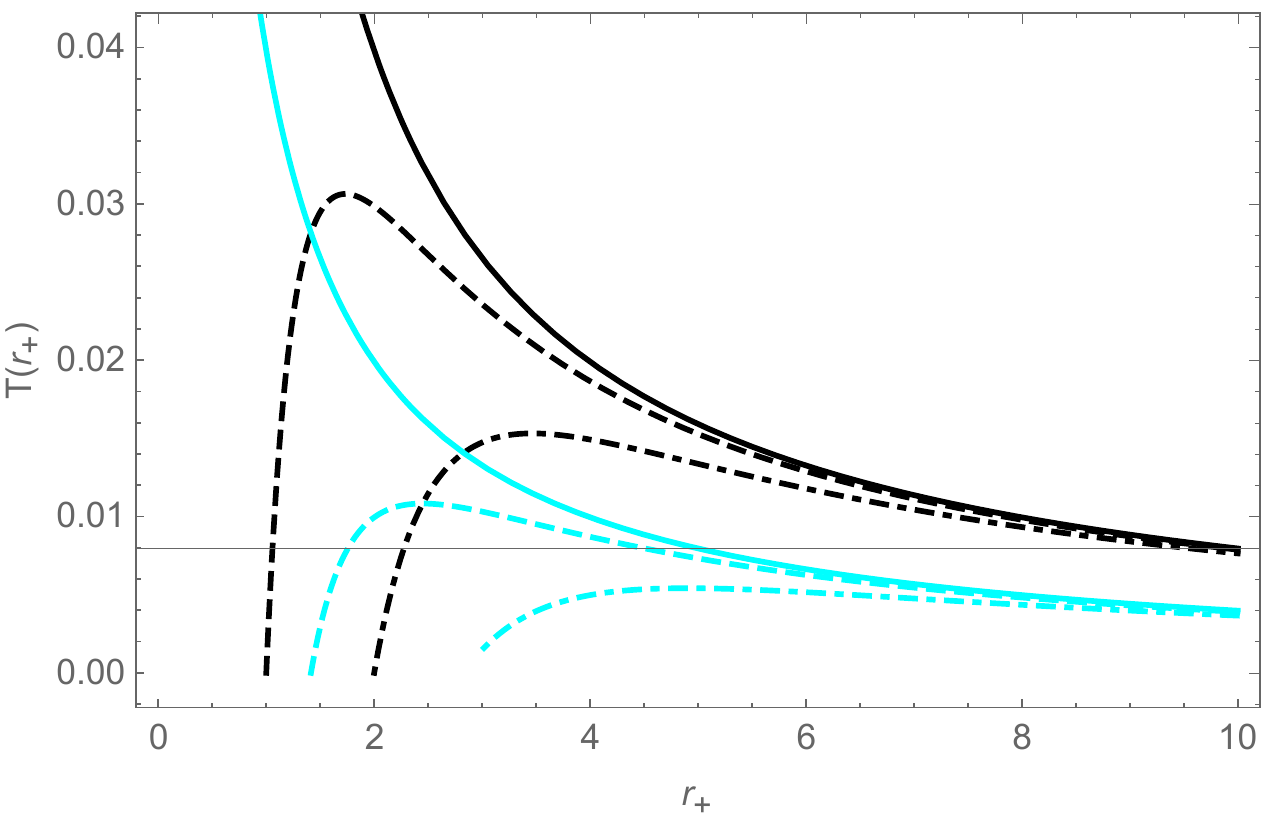} }
\subfigure[~$S(r_+)$]{
 \includegraphics[width=.45\textwidth]{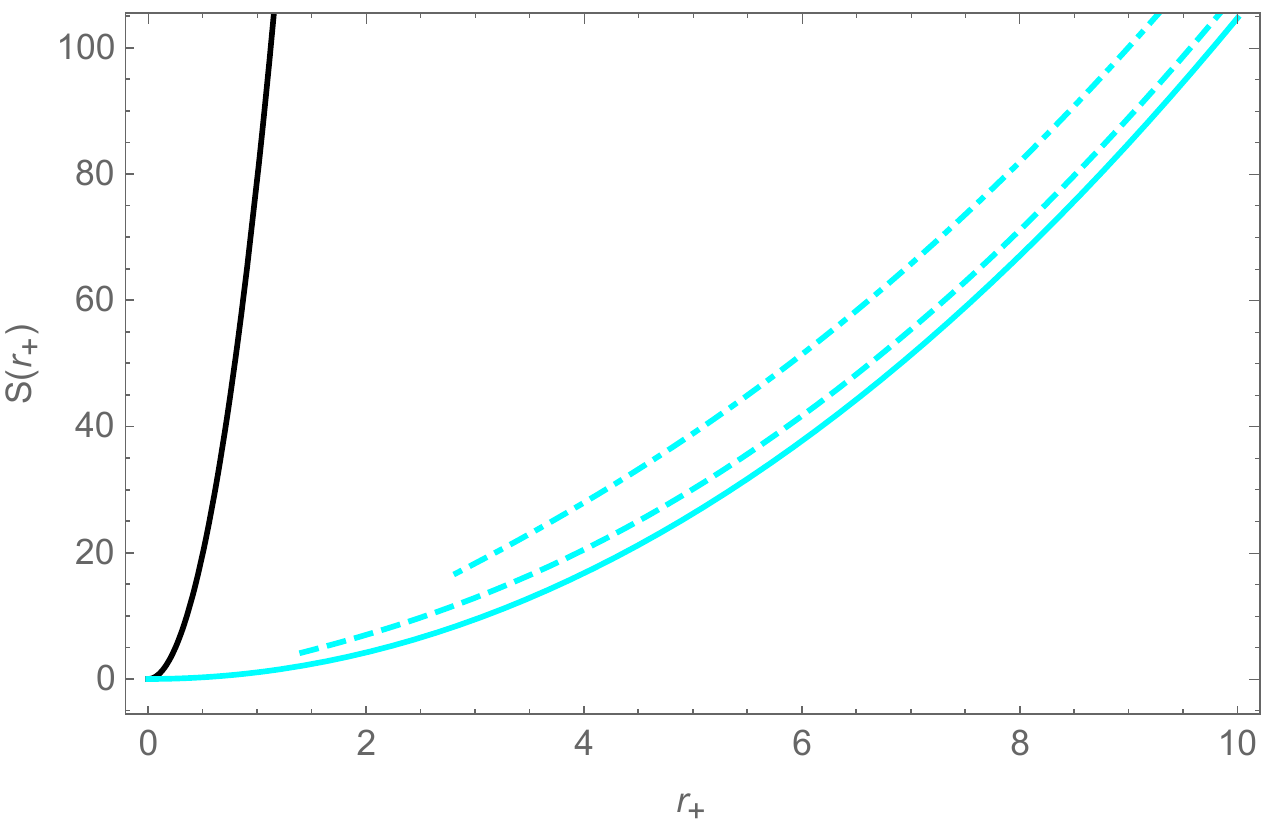} }
 \subfigure[~$M(m)$]{
 \includegraphics[width=.45\textwidth]{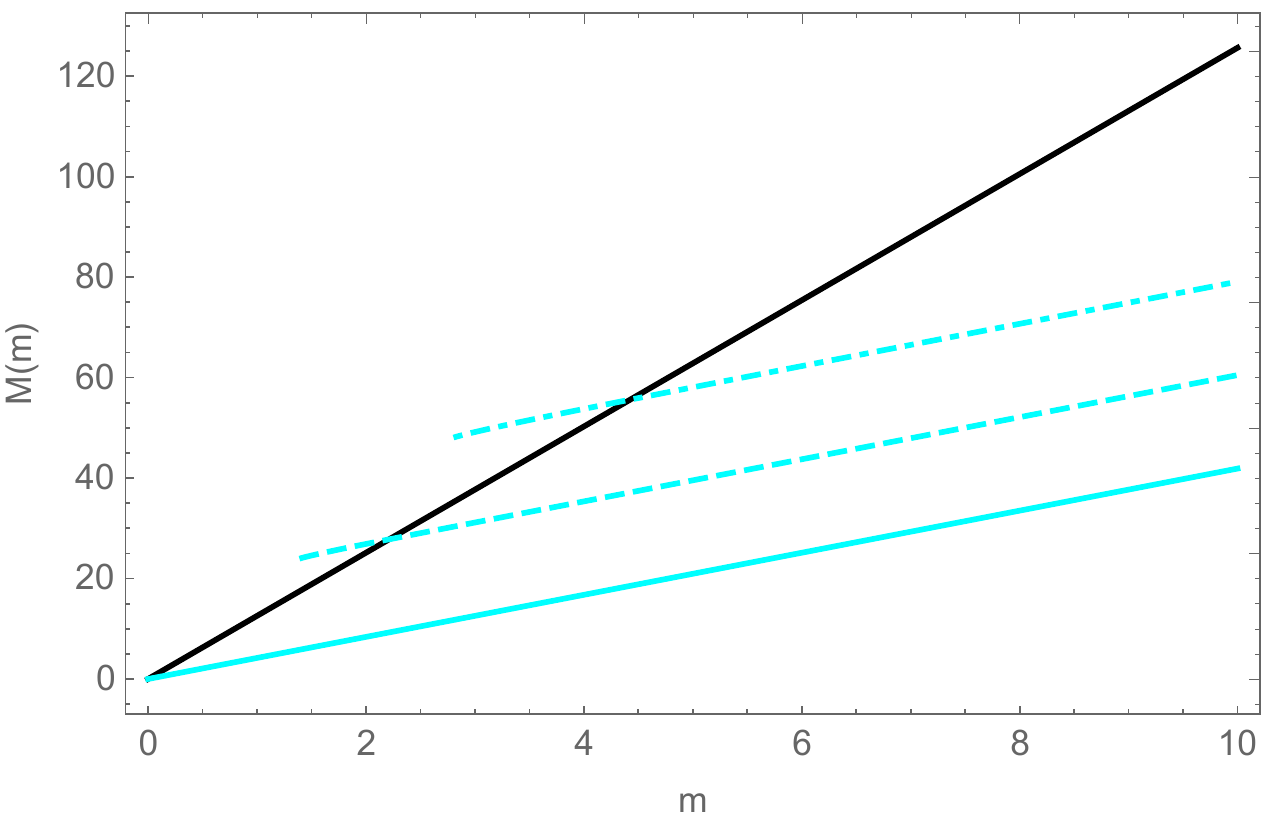} }
\subfigure[~$C_Q(r_+)$]{
 \includegraphics[width=.45\textwidth]{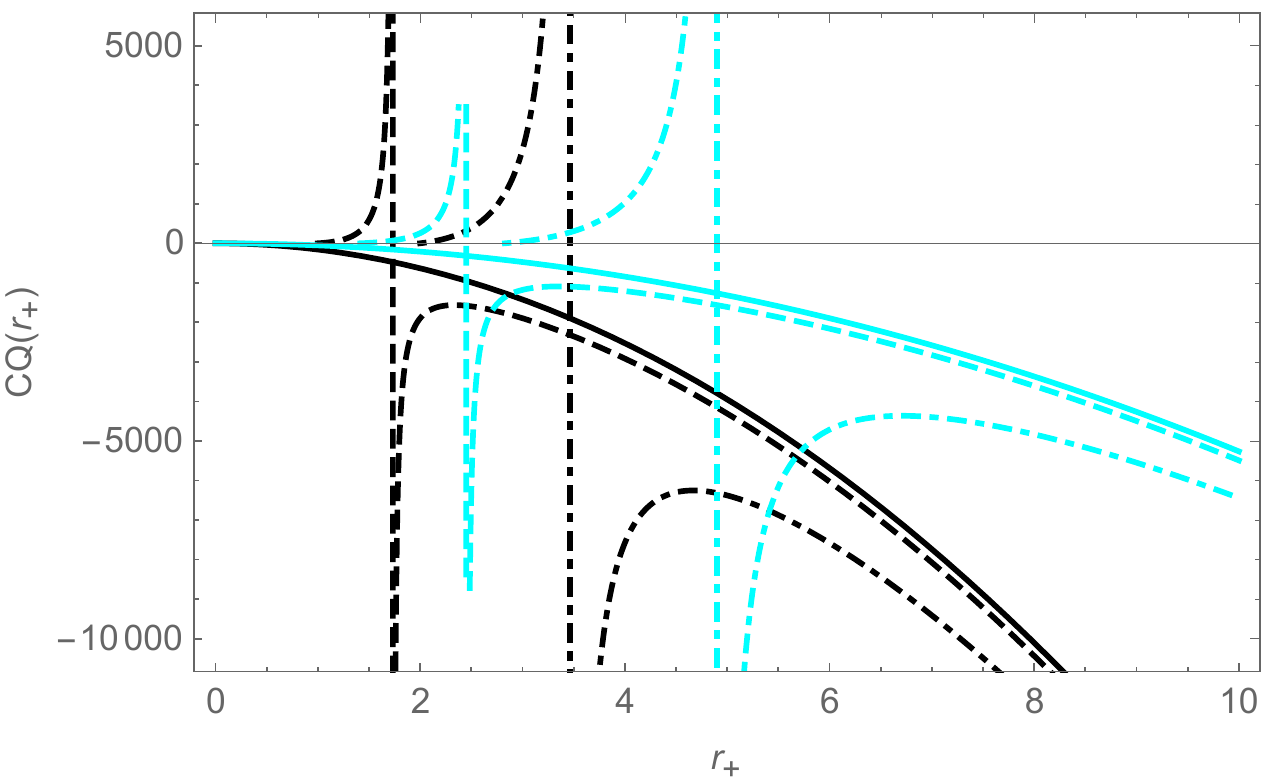} }
\caption{{} The black curves represent RN black hole in Einstein gravity while the cyan curves represent our black hole in $f(R)$ gravity in $(3+1)$-dimensions. In addition, the full lines, the dashed lines and the dotdashed lines correspond to $q=0,1,2$ respectively.}
\label{fig:TSMCQ4}
\end{figure}

Although the metric function looks similar with RN black hole, however the mass $M$ of the black hole can not be described only by the model parameter $m$. In other words, the real mass of the black hole is contributed by both model parameter $m$ and charge parameter $q$ due to the nonlinearity of $f(R)$. The expression (\ref{C_Q}) clearly shows that the heat capacity becomes zero at the extremal case $q=r_+/\sqrt{2}$ and divergent with a smaller charge $q=r_+/\sqrt{6}$, similar with the RN case in Einstein gravity. To compare with the RN black holes in Einstein gravity, we plot the thermodynamic quantities of both RN black hole in Einstein gravity and our black hole in $f(R)$ gravity in $(3+1)$-dimensions in FIG.\ref{fig:TSMCQ4}.

The Hawking temperature always decreases with the increase of the radius of the event horizon for the neutral case, indicating that neutral large black holes have lower Hawking temperature. While for the charged case the Hawking temperature goes up from zero (extremal case) to a maximum value then falls down to zero at infinity, where the electric charge can protect the black holes from the violent Hawking evaporation for small black holes. For black holes carrying the same charge, the black holes in $f(R)$ gravity always have lower temperatures than their Einstein cousins.

The Bekenstein-Hawking entropy in Einstein gravity is proportional to area, while in $f(R)$ gravity this relation is modified and leads to much lower entropy though charge can increase the entropy.

On the other side, the mass of RN black hole in Einstein gravity can be described linearly by a single parameter $m$, however in $f(R)$ gravity the real thermodynamic mass of the black hole are decided by both charge parameter $q$ and model parameter $m$ or $c_2$. In other words, this kind of $f(R)$ black hole can carry the information the model parameters. From the figure we can see that for fixed charge, the relation between $M$ and parameter $m$ is still asymptotically linear. We can check the derivative of $M(m)$
\begin{equation}
\lim_{q\to 0} M'(m)=\frac{4\pi}{3},  \quad  \lim_{m\to\infty} M'(m)=\frac{4\pi}{3}.
\end{equation}

Finally the heat capacity is always negative in neutral case for both Schwarzschild black hole and $f(R)$ black hole, which indicates that the absorption of energy cools the black hole while the emission of energy heats up the black hole and makes it unstable through Hawking radiation. In addition, the presence of charge makes heat capacity divergent at some critical points as we have shown and positive branches of heat capacity appear in front of the divergent points, meaning small black hole (small radius of the event horizon) has positive heat capacity like ordinary objects. For the $f(R)$ black hole, the divergence of heat capacity happens at black hole with larger radius of event horizon, and its heat capacity has smaller absolute value, which means that its ability to absorb and emit energy are both weaker than RN black hole.

\section{Conclusions}
\label{coclusion}

In this work we obtained an exact charged black hole solution with dynamic curvature in $D$-dimensions in Maxwell-$f(R)$ gravity. Without specifying the form of $f(R)$ we solved the Einstein-Maxwell equations under a metric ansatz with $g_{tt}g_{rr}=-1$. The general black hole solution we found, depending on the choices of the parameters, can reduce to the Reissner-Nordstr\"om  black hole in $D$-dimensions in Einstein gravity and to the known charged black hole solution with constant curvature in $f(R)$ gravity. We also obtained new charged $D$-dimensional solutions in the case of $c_1=-1$.  
All of our  general solutions in various dimensions consist of only polynomials and logarithmic terms, and with the constraints
they reveal interesting properties when compared to RN black holes.

In $(3+1)$-dimensions the form of $f(R)$ can be solved explicitly and the found polynomial solution with dynamic curvature was compared with the usual $(3+1)$-dimensions  RN black hole in the  Einstein gravity. The main characteristics of our solutions  are the presence of a dynamic curvature $R(r)$ and a non-zero gravitational action $F(r)$, while the RN black hole has a constant curvature and zero gravitational action. This fact gives a rich spectrum of thermodynamical properties of our charge black hole solution.

We studied analytically the thermodynamics of the charge black hole in $(3+1)$-dimensions and compared with the RN black holes in Einstein gravity, calculating the thermodynamical mass, Hawking temperature, Bekenstein-Hawking entropy and heat capacity. Firstly, the thermodynamical mass of the black hole we obtained via the First Law reveals the entanglement of the parameter $m$ and $q$ when they contribute to the black hole mass. In Einstein gravity, the parameter $m$ and $q$ represent the mass and charge respectively, while in the black hole solution in $f(R)$ gravity, the real mass is a complicated combination of parameter $m$ and charge $q$ due to the nonlinear of the $f(R)$ gravity. Then for the other thermodynamical quantities the behaviors of the black holes in $f(R)$ gravity are similar with the RN black holes in Einstein gravity in general, but the former ones possess lower Hawking temperature, lower Bekenstein-Hawking entropy and lower absolute value of the heat capacity.

\section{Acknowledgement}

We thank Yen Chin Ong, Martin Krssak, Dongchao Zheng, Mahdis Ghodrati and Athanasios Bakopoulos for extensive discussions. E.P. acknowledges the hospitality of the
Center for Gravitation and Cosmology, College of Physical Science
and Technology of the Yangzhou University, where part of this work was carried out.

\appendix
\section{Explicit solutions in $D=4,5,6$ dimensions}
\label{appendix1}

In this Appendix we list the solutions for the metric function in $D=4,5,6$ dimensions.

\begin{eqnarray}
B(r)&=&\frac{1}{12 \left(c_1+1\right){}^5 r^2}\left\{ 12 c_2^2 r^4 \left[\left(c_1+1\right){}^3 k+c_2 c_4 \left(c_1+1\right)+c_2^2 q^2\right] \ln \left|\frac{c_2 r+c_1+1}{r}\right| \right.\notag \\
&&+\left(c_1+1\right) \left[2 \left(c_1+1\right) r \left(3 \left(c_1+1\right) k r \left(3 c_2^2 r^2+c_1 \left(4-2 c_2 r\right)-2 c_2 r+2 c_1^2+2\right)\right.\right.\notag\\
&&\left.+6 c_3 \left(c_1+1\right){}^3 r^3-6 c_2^2 c_4 r^2+3 c_2 c_4 \left(c_1+1\right) r-2 c_4 \left(c_1+1\right){}^2\right)+q^2 \left(-12 c_2^3 r^3 \right.\notag\\
&&\left.\left.\left.+c_1 \left(6 c_2^2 r^2-8 c_2 r+9\right)+6 c_2^2 r^2+c_1^2 \left(9-4 c_2 r\right)-4 c_2 r+3 c_1^3+3\right)\right] \right\}  \quad D=4~, \label{General Solution4}\\
B(r)&=&\frac{1}{360 \left(c_1+1\right){}^7 r^4}\left\{60 c_2^2 r^6 \left(4 \left(c_1+1\right){}^5 k-6 c_2^2 c_4 \left(c_1+1\right){}^2+3 c_2^4 q^2\right) \ln\left|\frac{c_2 r+c_1+1}{r}\right|\right.\notag\\
&&+\left(c_1+1\right) \left(40 \left(c_1+1\right){}^4 k r^4 \left(5 c_2^2 r^2-6 c_1 \left(c_2 r-3\right)-6 c_2 r+9 c_1^2+9\right)+30 \left(c_1+1\right){}^2 r^2\right.\notag\\
&& \left(12 c_3 \left(c_1+1\right){}^4 r^4+12 c_2^3 c_4 r^3-6 c_2^2 c_4 \left(c_1+1\right) r^2+4 c_2 c_4 \left(c_1+1\right){}^2 r-3 c_4 \left(c_1+1\right){}^3\right)\notag\\
&&+3 q^2 \left(-60 c_2^5 r^5+30 c_2^4 r^4-20 c_2^3 r^3+c_1^3 \left(15 c_2^2 r^2-48 c_2 r+100\right)+15 c_2^2 r^2 \right.\notag\\
&&+c_1^2 \left(-20 c_2^3 r^3+45 c_2^2 r^2-72 c_2 r+100\right)+c_1 \left(30 c_2^4 r^4-40 c_2^3 r^3+45 c_2^2 r^2-48 c_2 r+50\right)\notag\\
&&\left.\left.\left.+c_1^4 \left(50-12 c_2 r\right)-12 c_2 r+10 c_1^5+10\right)\right)\right\}  \quad   D=5~, \label{General Solution5}\\
B(r)&=&\frac{1}{2520 \left(c_1+1\right){}^9 r^6}\left\{420 c_2^2 r^8 \left(3 \left(c_1+1\right){}^7 k+6 c_2^3 c_4 \left(c_1+1\right){}^3+2 c_2^6 q^2\right) \ln \left|\frac{c_2 r+c_1+1}{r}\right|\right.\notag\\
&&+420 c_2 r^7 \left(-3 \left(c_1+1\right){}^6 k-3 c_1 \left(c_1+1\right){}^6 k-6 c_2^3 c_4 \left(c_1+1\right){}^3-2 c_2^6 q^2\right)\notag\\
&&+420 r^6 \left(6 c_1^2 \left(c_1+1\right){}^6 k+6 \left(c_1+1\right){}^6 k+12 c_1 \left(c_1+1\right){}^6 k+3 c_2^3 c_4 \left(c_1+1\right){}^4+c_1 c_2^6 q^2+c_2^6 q^2\right)\notag\\
&&+\left(c_1+1\right) \left[r^8 \left(735 c_2^2 \left(c_1+1\right){}^6 k+2520 c_3 \left(c_1+1\right){}^8\right)-120 \left(c_1+1\right){}^6 c_2 q^2 r+105 \left(c_1+1\right){}^7 q^2\right.\notag\\
&&-280 \left(c_1+1\right){}^2 c_2^2 \left(3 c_4 \left(c_1+1\right){}^3+c_2^3 q^2\right) r^5+210 \left(c_1+1\right){}^3 c_2 \left(3 c_4 \left(c_1+1\right){}^3+c_2^3 q^2\right) r^4  \notag\\
&&\left.\left.-168 \left(c_1+1\right){}^4 \left(3 c_4 \left(c_1+1\right){}^3+c_2^3 q^2\right) r^3+140 \left(c_1+1\right){}^5 c_2^2 q^2 r^2\right]\right\}  \quad   D=6~, \label{General Solution6}
\end{eqnarray}
here $c_3$ and $c_4$ are new constants of integration, not the same with the constants of integration $c_3$ and $c_4$ of the general solution (\ref{General Solution}). This is because if we set $D=4,5,6...$, the general solution (\ref{General Solution}) will reduce to different solutions given by  (\ref{General Solution4}),(\ref{General Solution5}) and (\ref{General Solution6}) with different integration constants.

\section{Solutions with an effective cosmological constant}
\label{appendix2}

In this Appendix we give the solutions with an effective cosmological constant,
\begin{eqnarray}
B(r)&=&\frac{(D-3) k}{D-2}-\frac{2\Lambda_\text{eff}}{(D-1)(D-2)(c_1+1)}  r^2+\sum _{n=1}^{D-3} \frac{2 k (-1)^{n+1} (c_1+1)^n}{(D-2) (n+2) c_2^n r^n} \notag \\
&&+\sum _{n=D-2}^{2 (D-3)} \frac{q^2 (-1)^n c_2^{2 D-n-6}}{(D-3) (n+2) r^n (c_1+1)^{2 D-n-5}}~,\label{MetricD}\\
R(r)&=&\frac{(D-3) k}{r^2}+\frac{2 D}{(D-2)(c_1+1)}\Lambda_\text{eff}+\sum _{n=1}^{D-3} \frac{2 k (-1)^n (c_1+1)^n  \left(D^2-D (2 n+5)+n^2+5 n+6\right) }{(D-2) (n+2)c_2^{n} r^{n+2}} \notag \\
&&-\sum _{n=D-2}^{2 (D-3)} \frac{q^2 (-1)^n \left(D^2-D (2 n+5)+n^2+5 n+6\right)  }{(D-3) (n+2)(c_1+1)^{2 D-n-5} c_2^{n+6-2D}r^{n+2}}~,\\
f(r)&=&-\frac{2(D-2-2c_1)}{(D-2)(c_1+1)}\Lambda_\text{eff}+2\Lambda+\frac{2 c_2 (D-3) k}{r}+\frac{(D-3) k (2 c_1-D+4)}{(D-2) r^2}-\frac{q^2}{2 r^{2 D-4}} \notag\\
&&-\sum _{n=1}^{D-3} \frac{2 k (-1)^n (c_1+1)^n  (-D+n+3)  (2 c_1-D+n+4)}{(D-2) (n+2)c_2^{n} r^{n+2}}\notag\\
&&+\sum _{n=D-2}^{2 (D-3)} \frac{q^2 (-1)^n  (-D+n+3) (2c_1-D+n+4)}{(D-3) (n+2)(c_1+1)^{2 D-n-5} c_2^{n+6-2D}r^{n+2}}~,
\end{eqnarray}
and the equation of $f(R)$ also becomes simpler
\begin{eqnarray}
f(R)&=&-\frac{2(D-2-2c_1)}{(D-2)}\Lambda_\text{eff}+2\Lambda+\frac{2 c_2^2 (D-3) k}{f'(R)-c_1}+\frac{(D-3) k (2 c_1-D+4)c_2^2}{(D-2) \left(f'(R)-c_1\right)^2}-\frac{q^2 c_2^{2 D-4}}{2 \left(f'(R)-c_1\right)^{2 D-4}} \notag\\
&&-\sum _{n=1}^{D-3} \frac{2 k (-1)^n (c_1+1)^n c_2^{2} (-D+n+3)  (2 c_1-D+n+4)}{(D-2) (n+2) \left(f'(R)-c_1\right)^{n+2}}\notag\\
&&+\sum _{n=D-2}^{2 (D-3)} \frac{q^2 (-1)^n c_2^{2D-4} (-D+n+3) (2 c_1-D+n+4) }{(D-3) (n+2)(c_1+1)^{2 D-n-5}\left(f'(R)-c_1\right)^{n+2}}~.
\end{eqnarray}
Then the full gravity action becomes
\begin{eqnarray}
R(r)+f(r)-2\Lambda&=&\frac{4\Lambda_\text{eff}}{(D-2)}+\frac{2 c_2 (D-3) k}{r}+\frac{2 (c_1+1) (D-3) k}{(D-2) r^2}-\frac{q^2}{2 r^{2 D-4}}\notag \\
&& -\sum _{n=1}^{D-3} \frac{4 k (-1)^n (c_1+1)^{n+1} (-D+n+3) }{(D-2) (n+2)c_2^{n} r^{n+2}} \notag\\
&&+\sum _{n=D-2}^{2 (D-3)} \frac{2 q^2 (-1)^n (-D+n+3)  (c_1+1)^{-2 D+n+6} }{(D-3) (n+2)c_2^{n+6-2D}r^{n+2}}~.
\end{eqnarray}


\begin{thebibliography}{99}

\bibitem{Nojiri:2006ri}
  S.~Nojiri and S.~D.~Odintsov,
 ``Introduction to modified gravity and gravitational alternative for dark energy,''
  Conf C {\bf 0602061}, 06 (2006)
  [Int.\ J.\ Geom.\ Meth.\ Mod.\ Phys.\  {\bf 4}, 115 (2007)]
  [hep-th/0601213].

\bibitem{Copeland:2006wr}
  E.~J.~Copeland, M.~Sami and S.~Tsujikawa,
  ``Dynamics of dark energy,''
  Int.\ J.\ Mod.\ Phys.\ D {\bf 15}, 1753 (2006)
  [hep-th/0603057].

\bibitem{Nojiri:2010wj}
  S.~Nojiri and S.~D.~Odintsov,
  ``Unified cosmic history in modified gravity: from F(R) theory to Lorentz non-invariant models,''
  Phys.\ Rept.\  {\bf 505}, 59 (2011)
  [arXiv:1011.0544 [gr-qc]].

\bibitem{Clifton:2011jh}
  T.~Clifton, P.~G.~Ferreira, A.~Padilla and C.~Skordis,
  ``Modified Gravity and Cosmology,''
  Phys.\ Rept.\  {\bf 513}, 1 (2012)
  [arXiv:1106.2476 [astro-ph.CO]].

\bibitem{Stelle:1976gc}
  K.~S.~Stelle,
  ``Renormalization of Higher Derivative Quantum Gravity,''
  Phys.\ Rev.\ D {\bf 16}, 953 (1977).

\bibitem{Gregory:2009fj}
R.~Gregory, S.~Kanno and J.~Soda,
``Holographic Superconductors with Higher Curvature Corrections,''
JHEP \textbf{10} (2009), 010
[arXiv:0907.3203 [hep-th]].

\bibitem{Kuang:2013oqa}
X.~M.~Kuang, E.~Papantonopoulos, G.~Siopsis and B.~Wang,
``Building a Holographic Superconductor with Higher-derivative Couplings,''
Phys. Rev. D \textbf{88}, 086008 (2013)
[arXiv:1303.2575 [hep-th]].

\bibitem{Gasperini:1997up}
M.~Gasperini,
``Tensor perturbations in high curvature string backgrounds,''
Phys. Rev. D \textbf{56} (1997), 4815-4823
[arXiv:gr-qc/9704045 [gr-qc]].

\bibitem{Hansraj:2020xmz}
S.~Hansraj, M.~Govender, L.~Moodly and K.~N.~Singh,
``Influence of quadratic curvature invariants on the dynamics of stellar structure,''
[arXiv:2003.04568 [gr-qc]].


\bibitem{DeFelice:2010aj}
  A.~De Felice and S.~Tsujikawa,
``f(R) theories,''
  Living Rev.\ Rel.\  {\bf 13}, 3 (2010),
  [arXiv:1002.4928 [gr-qc]].

\bibitem{Cognola:2007zu}
  G.~Cognola, E.~Elizalde, S.~Nojiri, S.~D.~Odintsov, L.~Sebastiani and S.~Zerbini,
  ``A Class of viable modified f(R) gravities describing inflation and the onset of accelerated expansion,''
  Phys.\ Rev.\ D {\bf 77} (2008) 046009
  [arXiv:0712.4017 [hep-th]].

\bibitem{Pogosian:2007sw}
  L.~Pogosian and A.~Silvestri,
  ``The pattern of growth in viable f(R) cosmologies,''
  Phys.\ Rev.\ D {\bf 77} (2008) 023503
   Erratum: [Phys.\ Rev.\ D {\bf 81} (2010) 049901]
  [arXiv:0709.0296 [astro-ph]].

\bibitem{Zhang:2005vt}
  P.~Zhang,
  ``Testing $f(R)$ gravity against the large scale structure of the universe.,''
  Phys.\ Rev.\ D {\bf 73} (2006) 123504
  [astro-ph/0511218].

\bibitem{Li:2007xn}
  B.~Li and J.~D.~Barrow,
  ``The Cosmology of f(R) gravity in metric variational approach,''
  Phys.\ Rev.\ D {\bf 75} (2007) 084010
  [gr-qc/0701111].

\bibitem{Song:2007da}
  Y.~S.~Song, H.~Peiris and W.~Hu,
  ``Cosmological Constraints on f(R) Acceleration Models,''
  Phys.\ Rev.\ D {\bf 76} (2007) 063517
  [arXiv:0706.2399 [astro-ph]].



\bibitem{Nojiri:2007cq}
  S.~Nojiri and S.~D.~Odintsov,
  ``Modified f(R) gravity unifying R**m inflation with Lambda CDM epoch,''
  Phys.\ Rev.\ D {\bf 77} (2008) 026007
  [arXiv:0710.1738 [hep-th]].

\bibitem{Nojiri:2007as}
  S.~Nojiri and S.~D.~Odintsov,
  ``Unifying inflation with LambdaCDM epoch in modified f(R) gravity consistent with Solar System tests,''
  Phys.\ Lett.\ B {\bf 657} (2007) 238
  [arXiv:0707.1941 [hep-th]].

\bibitem{Capozziello:2018ddp}
  S.~Capozziello, C.~A.~Mantica and L.~G.~Molinari,
  ``Cosmological perfect-fluids in f(R) gravity,''
  Int.\ J.\ Geom.\ Meth.\ Mod.\ Phys.\  {\bf 16} (2018) no.01,  1950008
  [arXiv:1810.03204 [gr-qc]].

\bibitem{Vainio:2016qas}
  J.~Vainio and I.~Vilja,
  ``$f(R)$ gravity constraints from gravitational waves,''
  Gen.\ Rel.\ Grav.\  {\bf 49} (2017) no.8,  99
  [arXiv:1603.09551 [astro-ph.CO]].

\bibitem{Ostrogradsky:1850fid}
  M.~Ostrogradsky,
  ``Mémoires sur les équations différentielles, relatives au problème des isopérimètres,''
  Mem.\ Acad.\ St.\ Petersbourg {\bf 6} (1850) no.4,  385.

\bibitem{Woodard:2006nt}
  R.~P.~Woodard,
  ``Avoiding dark energy with 1/r modifications of gravity,''
  Lect.\ Notes Phys.\  {\bf 720} (2007) 403
  [astro-ph/0601672].

\bibitem{Hendi:2009sw}
S.~H.~Hendi,
``The Relation between F(R) gravity and Einstein-conformally invariant Maxwell source,''
Phys. Lett. B \textbf{690} (2010), 220-223
[arXiv:0907.2520 [gr-qc]].

\bibitem{HabibMazharimousavi:2011yj}
S.~Habib Mazharimousavi, M.~Halilsoy and T.~Tahamtan,
``Solutions for f(R) gravity coupled with electromagnetic field,''
Eur. Phys. J. C \textbf{72} (2012), 1851
[arXiv:1110.5085 [gr-qc]].

\bibitem{Konoplya:2008au}
R.~A.~Konoplya and A.~Zhidenko,
``Instability of higher dimensional charged black holes in the de-Sitter world,''
Phys. Rev. Lett. \textbf{103} (2009), 161101
[arXiv:0809.2822 [hep-th]].

\bibitem{Cardoso:2010rz}
V.~Cardoso, M.~Lemos and M.~Marques,
``On the instability of Reissner-Nordstrom black holes in de Sitter backgrounds,''
Phys. Rev. D \textbf{80} (2009), 127502
[arXiv:1001.0019 [gr-qc]].

\bibitem{Konoplya:2008ix}
R.~A.~Konoplya and A.~Zhidenko,
``(In)stability of D-dimensional black holes in Gauss-Bonnet theory,''
Phys. Rev. D \textbf{77} (2008), 104004
[arXiv:0802.0267 [hep-th]].

\bibitem{Beroiz:2007gp}
M.~Beroiz, G.~Dotti and R.~J.~Gleiser,
``Gravitational instability of static spherically symmetric Einstein-Gauss-Bonnet black holes in five and six dimensions,''
Phys. Rev. D \textbf{76} (2007), 024012
[arXiv:hep-th/0703074 [hep-th]].

\bibitem{Gregory:1993vy}
R.~Gregory and R.~Laflamme,
``Black strings and p-branes are unstable,''
Phys. Rev. Lett. \textbf{70} (1993), 2837-2840
[arXiv:hep-th/9301052 [hep-th]].

\bibitem{Park:2005vw}
D.~K.~Park,
``Hawking radiation of the brane-localized graviton from a (4+n)-dimensional black hole,''
Class. Quant. Grav. \textbf{23} (2006), 4101-4110
[arXiv:hep-th/0512021 [hep-th]].

\bibitem{Mkenyeleye:2015jta}
M.~D.~Mkenyeleye, R.~Goswami and S.~D.~Maharaj,
``Is cosmic censorship restored in higher dimensions?,''
Phys. Rev. D \textbf{92} (2015) no.2, 024041
[arXiv:1503.06651 [gr-qc]].

\bibitem{Liu:2019lon}
H.~Liu, Z.~Tang, K.~Destounis, B.~Wang, E.~Papantonopoulos and H.~Zhang,
``Strong Cosmic Censorship in higher-dimensional Reissner-Nordström-de Sitter spacetime,''
JHEP \textbf{03} (2019), 187
[arXiv:1902.01865 [gr-qc]].

\bibitem{Herdeiro:2013pia}
C.~A.~R.~Herdeiro, J.~C.~Degollado and H.~F.~Rúnarsson,
``Rapid growth of superradiant instabilities for charged black holes in a cavity,''
Phys. Rev. D \textbf{88} (2013), 063003
[arXiv:1305.5513 [gr-qc]].

\bibitem{Hod:2013fvl}
S.~Hod,
``Analytic treatment of the charged black-hole-mirror bomb in the highly explosive regime,''
Phys. Rev. D \textbf{88} (2013) no.6, 064055
[arXiv:1310.6101 [gr-qc]].

\bibitem{Zhang:2013haa}
S.~J.~Zhang, B.~Wang and E.~Abdalla,
``Superradiant instability of extremal brane-world Reissner-Nordstr\"{o}m black holes to charged scalar perturbations,''
[arXiv:1306.0932 [gr-qc]].

\bibitem{Zhu:2014sya}
Z.~Zhu, S.~J.~Zhang, C.~E.~Pellicer, B.~Wang and E.~Abdalla,
``Stability of Reissner-Nordström black hole in de Sitter background under charged scalar perturbation,''
Phys. Rev. D \textbf{90} (2014) no.4, 044042
[arXiv:1405.4931 [hep-th]].


\bibitem{Myers:1986un}
R.~C.~Myers and M.~J.~Perry,
``Black Holes in Higher Dimensional Space-Times,''
Annals Phys. \textbf{172} (1986), 304


\bibitem{Multamaki:2006zb}
  T.~Multamaki and I.~Vilja,
  ``Spherically symmetric solutions of modified field equations in f(R) theories of gravity,''
  Phys.\ Rev.\ D {\bf 74} (2006) 064022
  [astro-ph/0606373].

\bibitem{Sebastiani:2010kv}
  L.~Sebastiani and S.~Zerbini,
  ``Static Spherically Symmetric Solutions in F(R) Gravity,''
  Eur.\ Phys.\ J.\ C {\bf 71} (2011) 1591
  [arXiv:1012.5230 [gr-qc]].

\bibitem{Amirabi:2015aya}
Z.~Amirabi, M.~Halilsoy and S.~Habib Mazharimousavi,
``Generation of spherically symmetric metrics in f(R) gravity,''
Eur. Phys. J. C \textbf{76} (2016) no.6, 338
[arXiv:1509.06967 [gr-qc]].

\bibitem{delaCruzDombriz:2009et}
  A.~de la Cruz-Dombriz, A.~Dobado and A.~L.~Maroto,
  ``Black Holes in f(R) theories,''
  Phys.\ Rev.\ D {\bf 80} (2009) 124011
   Erratum: [Phys.\ Rev.\ D {\bf 83} (2011) 029903]
  [arXiv:0907.3872 [gr-qc]].

\bibitem{Moon:2011hq}
  T.~Moon, Y.~S.~Myung and E.~J.~Son,
  ``f(R) black holes,''
  Gen.\ Rel.\ Grav.\  {\bf 43} (2011) 3079
  [arXiv:1101.1153 [gr-qc]].

\bibitem{Cembranos:2011sr}
  J.~A.~R.~Cembranos, A.~de la Cruz-Dombriz and P.~Jimeno Romero,
  ``Kerr-Newman black holes in $f(R)$ theories,''
  Int.\ J.\ Geom.\ Meth.\ Mod.\ Phys.\  {\bf 11} (2014) 1450001
  [arXiv:1109.4519 [gr-qc]].

\bibitem{Multamaki:2006ym}
  T.~Multamaki and I.~Vilja,
  ``Static spherically symmetric perfect fluid solutions in f(R) theories of gravity,''
  Phys.\ Rev.\ D {\bf 76} (2007) 064021
  [astro-ph/0612775].

\bibitem{Mazharimousavi:2011bf}
S.~H.~Mazharimousavi, M.~Halilsoy and T.~Tahamtan,
``Constant curvature f(R) gravity minimally coupled with Yang-Mills field,''
Eur. Phys. J. C \textbf{72} (2012), 1958
[arXiv:1109.3655 [gr-qc]].

\bibitem{Mazharimousavi:2011nc}
S.~H.~Mazharimousavi and M.~Halilsoy,
``Black hole solutions in f(R) gravity coupled with non-linear Yang-Mills field,''
Phys. Rev. D \textbf{84} (2011), 064032
[arXiv:1105.3659 [gr-qc]].

\bibitem{Hollenstein:2008hp}
  L.~Hollenstein and F.~S.~N.~Lobo,
  ``Exact solutions of f(R) gravity coupled to nonlinear electrodynamics,''
  Phys.\ Rev.\ D {\bf 78} (2008) 124007
  [arXiv:0807.2325 [gr-qc]].

\bibitem{Rodrigues:2015ayd}
M.~E.~Rodrigues, E.~L.~Junior, G.~T.~Marques and V.~T.~Zanchin,
``Regular black holes in $f(R)$ gravity coupled to nonlinear electrodynamics,''
Phys. Rev. D \textbf{94} (2016) no.2, 024062
[arXiv:1511.00569 [gr-qc]].

\bibitem{Hurtado:2020gic}
R.~A.~Hurtado and J.~R.~Arenas,
``Spherically symmetric and static solutions in $f(R)$ gravity coupled with EM fields,''
[arXiv:2002.06059 [gr-qc]].

\bibitem{Capozziello:2009jg}
  S.~Capozziello, M.~De laurentis and A.~Stabile,
  ``Axially symmetric solutions in f(R)-gravity,''
  Class.\ Quant.\ Grav.\  {\bf 27} (2010) 165008
  [arXiv:0912.5286 [gr-qc]].

\bibitem{Hendi:2011eg}
  S.~H.~Hendi, B.~Eslam Panah and S.~M.~Mousavi,
  ``Some exact solutions of F(R) gravity with charged (a)dS black hole interpretation,''
  Gen.\ Rel.\ Grav.\  {\bf 44}, 835 (2012)
  [arXiv:1102.0089 [hep-th]].


\bibitem{Amirabi:2015aya}
Z.~Amirabi, M.~Halilsoy and S.~Habib Mazharimousavi,
``Generation of spherically symmetric metrics in f(R) gravity,''
Eur. Phys. J. C \textbf{76} (2016) no.6, 338
[arXiv:1509.06967 [gr-qc]].

\bibitem{Carames:2009ek}
T.~R.~P.~Carames and E.~R.~Bezerra de Mello,
``Spherically symmetric vacuum solutions of modified gravity theory in higher dimensions,''
Eur. Phys. J. C \textbf{64} (2009), 113-121
[arXiv:0901.0814 [gr-qc]].


\bibitem{Ong:2015fha}
  Y.~C.~Ong,
  ``Hawking Evaporation Time Scale of Topological Black Holes in Anti-de Sitter Spacetime,''
  Nucl.\ Phys.\ B {\bf 903} (2016) 387
  [arXiv:1507.07845 [gr-qc]].

\bibitem{Hendi:2014wsa}
  S.~H.~Hendi,
  ``(2+1)-Dimensional Solutions in $F(R)$ Gravity,''
  Int.\ J.\ Theor.\ Phys.\  {\bf 53}, no. 12, 4170 (2014)
  [arXiv:1410.7527 [gr-qc]].


\bibitem{Chabab:2016cem}
  M.~Chabab, H.~El Moumni, S.~Iraoui and K.~Masmar,
  ``Behavior of quasinormal modes and high dimension RN–AdS black hole phase transition,''
  Eur.\ Phys.\ J.\ C {\bf 76} (2016) no.12,  676
  [arXiv:1606.08524 [hep-th]].



\bibitem{Banados:1992wn}
  M.~Banados, C.~Teitelboim and J.~Zanelli,
  ``The Black hole in three-dimensional space-time,''
  Phys.\ Rev.\ Lett.\  {\bf 69} (1992) 1849
  [hep-th/9204099].











%
%
%
%
%
%
%
%
%
%
%
%
%
%
%
%
%
%
%
%
%
%
%
%
%
%
%
%
%
%
%
%
%
%
%
%
%
%
%
%







\bibitem{Soroushfar:2016nbu}
  S.~Soroushfar, R.~Saffari and N.~Kamvar,
  ``Thermodynamic geometry of black holes in f(R) gravity,''
  Eur.\ Phys.\ J.\ C {\bf 76} (2016) no.9,  476
  [arXiv:1605.00767 [gr-qc]].



\bibitem{Nashed:2019tuk}
  G.~G.~L.~Nashed and S.~Capozziello,
 ``Charged spherically symmetric black holes in $f(R)$ gravity and their stability analysis,''
  Phys.\ Rev.\ D {\bf 99}, no. 10, 104018 (2019)
  [arXiv:1902.06783 [gr-qc]].

\bibitem{Akbar:2006mq}
M.~Akbar and R.~G.~Cai,
``Thermodynamic Behavior of Field Equations for f(R) Gravity,''
Phys. Lett. B \textbf{648}, 243-248 (2007)
[arXiv:gr-qc/0612089 [gr-qc]].


\bibitem{Camci:2020yre}
U.~Camci,
``Three-dimensional black holes via Noether symmetries,''
[arXiv:2012.06064 [gr-qc]].




%
%

\bibitem{Cognola:2011nj}
  G.~Cognola, O.~Gorbunova, L.~Sebastiani and S.~Zerbini,
 ``On the Energy Issue for a Class of Modified Higher Order Gravity Black Hole Solutions,''
  Phys.\ Rev.\ D {\bf 84}, 023515 (2011)
  [arXiv:1104.2814 [gr-qc]].

\bibitem{Zheng:2018fyn}
  Y.~Zheng and R.~J.~Yang,
  ``Horizon thermodynamics in $f(R)$ theory,''
  Eur.\ Phys.\ J.\ C {\bf 78}, no. 8, 682 (2018)
  [arXiv:1806.09858 [gr-qc]].





\bibitem{Nouicer:2007pu}
  K.~Nouicer,
  ``Black holes thermodynamics to all order in the Planck length in extra dimensions,''
  Class.\ Quant.\ Grav.\  {\bf 24} (2007) 5917
   Erratum: [Class.\ Quant.\ Grav.\  {\bf 24} (2007) 6435]
  [arXiv:0706.2749 [gr-qc]].


\bibitem{Frolov:2013efa}
  V.~P.~Frolov,
  ``Black Holes, Cosmology and Extra Dimensions,''
  Class.\ Quant.\ Grav.\  {\bf 30} (2013) 199001.

\bibitem{Herrera:2014dga}
  R.~Herrera and N.~Videla,
  ``The generalized second law of thermodynamics for interacting $f(R)$ gravity,''
  Int.\ J.\ Mod.\ Phys.\ D {\bf 23} (2014) no.08,  1450071
  [arXiv:1406.6305 [gr-qc]].


\bibitem{Bekenstein:1973ur}
  J.~D.~Bekenstein,
  ``Black holes and entropy,''
  Phys.\ Rev.\ D {\bf 7} (1973) 2333.

\bibitem{He:2019kws}
  K.~J.~He, G.~P.~Li and X.~Y.~Hu,
  ``Violations of the weak cosmic censorship conjecture in the higher dimensional f(R) black holes with pressure,''
  arXiv:1909.09956 [hep-th].

\bibitem{Asadzadeh:2015zxa}
  S.~Asadzadeh, M.~S.~Khaledian and K.~Karami,
  ``Structure formation and generalized second law of thermodynamics in some viable $f(R)$-gravity models,''
  Astron.\ Astrophys.\  {\bf 3} (2016) 81
  [arXiv:1507.08538 [gr-qc]].

\end{thebibliography}
\end{document}